\documentclass[preprint2]{aastex}

\newcommand{\mqd}{M\,82}
\newcommand{\etal}{et al.}
\newcommand{\brg}{Br$\gamma$}
\newcommand{\ewcoa}{$W_{2.29}$}
\newcommand{\ewcoh}{$W_{1.62}$}
\newcommand{\masssol}{$\rm ~M_{\odot}$}
\newcommand{\teff}{$T_{\rm eff}$}
\newcommand{\mlow}{$m_{\rm low}$}
\newcommand{\mup}{$m_{\rm up}$}
\newcommand{\tsc}{$t_{\rm sc}$}
\newcommand{\tb}{$t_{\rm b}$}
\newcommand{\lbol}{$L_{\rm bol}$}
\newcommand{\lbolOB}{$L_{\rm bol}^{\rm OB}$}
\newcommand{\llyc}{$L_{\rm Lyc}$}
\newcommand{\lk}{$L_{K}$}
\newcommand{\mlk}{$M^{\star}/L_{K}$}

\newcommand{\mstars}{$M^{\star}$}
\newcommand{\snrate}{$\nu_{\rm SN}$}
\newcommand{\snrlbol}{$\nu_{\rm SN}/L_{\rm bol}$}
\newcommand{\massstb}{$M_{\rm stb}^{\star}$}
\newcommand{\neonteff}
           {[\ion{Ne}{3}] 15.6\,$\rm \mu m$ / [\ion{Ne}{2}] 12.8$\,\rm \mu m$}

\slugcomment{Accepted for publication in Astrophysical Journal}

\shorttitle{The Nature of Starburst Activity in M\,82}
\shortauthors{F\"orster Schreiber \etal}

\begin{document}

\title{The Nature of Starburst Activity in M\,82
   \footnote{Based on observations with {\em ISO\/}, an ESA project with
   instruments funded by ESA Member States (especially the PI countries:
   France, Germany, the Netherlands, and the United Kingdom) and with the
   participation of ISAS and NASA. The SWS is a joint project of SRON and MPE.}
}

\author{N. M. F\"orster Schreiber\altaffilmark{1}, R. Genzel, D. Lutz}
\affil{Max-Planck-Institut f\"ur Extraterrestrische Physik, 
       Postfach 1312, D-85741 Garching, Germany}
\email{forster@strw.leidenuniv.nl, genzel@mpe.mpg.de, lutz@mpe.mpg.de}
\and 
\author{A. Sternberg}
\affil{School of Physics and Astronomy, Tel Aviv University, 
       Ramat Aviv, Tel Aviv 69978, Israel}
\email{amiel@wise.tau.ac.il}
\altaffiltext{1}{Current address: Leiden Observatory, 
                 PO Box 9513, 2300 RA Leiden, The Netherlands}

\begin{abstract}

We present new evolutionary synthesis models of \mqd\ based mainly on
observations consisting of near-infrared integral field spectroscopy
and mid-infrared spectroscopy.  The models incorporate stellar evolution, 
spectral synthesis, and photoionization modeling, and are optimized for 
$\rm \lambda = 1 - 45~\mu m$ observations of starburst galaxies.
The data allow us to model the starburst regions on scales as small as 25~pc. 
We investigate the initial mass function (IMF) of the stars and constrain 
quantitatively the spatial and temporal evolution of starburst activity
in \mqd.
We find a typical decay timescale for individual burst sites of a few million
years.  The data are consistent with the formation of very massive stars
($\rm \gtrsim 50 - 100~M_{\odot}$) and require a flattening of the starburst
IMF below a few solar masses assuming a Salpeter slope
${\rm d}N/{\rm d}m \propto m^{-2.35}$ at higher masses.
Our results are well matched by a scenario in which the global starburst
activity in \mqd\ occurred in two successive episodes each lasting a few
million years, peaking about $\rm 10^7~yr$ and $\rm 5 \times 10^{6}~yr$ ago.
The first episode took place throughout the central regions of \mqd\ and
was particularly intense at the nucleus while the second episode occurred
predominantly in a circumnuclear ring and along the stellar bar.  
We interpret this sequence as resulting from the gravitational interaction
between \mqd\ and its neighbour M\,81, and subsequent bar-driven evolution.
The short burst duration on all spatial scales indicates strong negative
feedback effects of starburst activity, both locally and globally.
Simple energetics considerations suggest the collective mechanical energy 
released by massive stars was able to rapidly inhibit star formation after
the onset of each episode.

\end{abstract}

\keywords{galaxies: evolution --- galaxies: individual (\mqd) ---
galaxies: starburst --- infrared: galaxies --- stars: formation}

\section{INTRODUCTION} \label{Intro}

In recent years, it has become clear that starburst galaxies are
important constituents of the Universe at all accessible redshifts
\citep*[e.g.,][]{Hec98, Ade00, Gia02, Sha01, Sha03, Cha03}.
However, despite extensive studies in the past two decades, a detailed
and quantitative understanding of the starburst phenomenon is still
lacking.  Crucial issues that remain open include the evolution and
feedback effects of starburst activity, its triggering and quenching
mechanisms, and the mass distribution of the stars formed in starbursts.
Progress has been hindered by the scarcity of spatially resolved data.
Furthermore, high resolution optical and ultraviolet studies are often
hampered by severe dust obscuration.

In this context, we have undertaken a study of the archetypal starburst
galaxy \mqd\ based on near-infrared (near-IR) integral field spectroscopy 
and mid-infrared (mid-IR) spectroscopy.  Our observations provide detailed
information on key features tracing various components of the interstellar
medium and of the stellar population, deep into the obscured star-forming
regions.
The data, along with results of nebular analysis and population synthesis,
are described by \citet[][hereafter paper~I]{FS01}, laying the observational
basis essential for the starburst modeling presented in this paper.
In a closely related study, \citet{Tho00} modeled the mid-IR line
emission of 27 starburst galaxies to examine the issues of massive star
formation and evolution in starburst environments.  Their results are
particularly relevant to our modeling of \mqd.

By virtue of its proximity and brightness, \mqd\ is an ideal target for
investigations of the starburst phenomenon.  Following the seminal paper
by \citet{Rie80}, several authors have applied evolutionary synthesis models
to \mqd\ \citep*[][among others]{Ber92, Rie93, Doa93, Sat97, Col99}.
However, a number of issues remain controversial.
For instance, it is still debated whether the initial mass function (IMF)
is biased against low-mass stars (see in particular \citealt{Rie93} and
\citealt{Sat97}).  It has also been suggested that the formation of stars
with masses $\rm \gtrsim 30~M_{\odot}$ may be suppressed 
\citep[e.g.,][]{Pux89}.  Furthermore, in analyzing the relative
distributions of various gaseous and stellar components, some authors
have proposed inside-out propagation of starburst activity in \mqd\
\citep[e.g.,][]{McL93, Sat97} while others have argued in favour
of the opposite scenario \citep*[e.g.,][]{She95, Gri00, Gri01}. 

The previous models of \mqd\ were either optimized to best reproduce the
global properties or focussed on selected stellar clusters.  Given the
complexity of \mqd, a full understanding requires spatially detailed
information of various diagnostics tracing different star formation epochs.
Our new IR data, complemented with results at other wavelengths from
the literature, allow us to probe the star formation history in the central
regions of \mqd\ ``continuously'' in space on scales as small as 25~pc and
in time up to at least $\rm \sim 50~Myr$ ago.  With the application of
starburst models we have developed, we use these data to re-examine the
issue of the IMF and constrain the spatial and temporal evolution of
the star formation activity.  Our model results provide quantitative
constraints for the triggering mechanisms and feedback effects of
starburst activity in \mqd.

The paper is organized as follows. 
Section~\ref{Sect-Obs} summarizes the observational constraints.
Section~\ref{Sect-models} describes the starburst models which are
applied to selected regions in \S~\ref{Sect-models_regions}.
Section~\ref{Sect-models_pixels} presents the spatially detailed modeling.
Section~\ref{Sect-discussion} discusses the results and
\S~\ref{Sect-conclu} summarizes the paper.

\section{OBSERVATIONAL CONSTRAINTS} \label{Sect-Obs}

Our modeling of \mqd\ uses the observations presented and analyzed in
paper~I.  The data consist of near-IR $H$- and $K$-band integral field
spectroscopy at $R \equiv \lambda/\Delta\lambda \sim 1000$ obtained with
the Max-Planck-Institut f\"ur extraterrestrische Physik (MPE)
3D instrument \citep{Wei96} and of $\rm \lambda = 2.4 - 45~\mu m$ 
spectroscopy at $R \sim 500 - 2000$ from the Short Wavelength Spectrometer 
(SWS; \citealt{deG96}) on board the {\em Infrared Space Observatory\/}
({\em ISO\/}; \citealt{Kes96}).  The 3D field of view is nearly parallel
to the galactic plane, includes the nucleus, and extends to the west out to
about 200~pc.  The SWS apertures cover regions including and centered
on the 3D field of view.  The 3D data have a spatial resolution of
$1.5^{\prime\prime}$, corresponding to about 25~pc at the distance of
\mqd\ (3.3~Mpc; \citealt{Fre88}).  We complemented these data with
results from the literature, providing further essential constraints.

We selected five representative regions for a detailed analysis.  These 
include the ``starburst core'' of \mqd\ corresponding to a 500\,pc--diameter
region centered on the nucleus, the 3D field of view covering the most
intense starburst regions, the central 35~pc at the nucleus, and two 
$\rm 35 \times 35~pc$ regions positioned at the brightest \brg\
sources observed with 3D, which we designated M82:Br1 and M82:Br2 in paper~I
($\approx 10^{\prime\prime}$ and $5^{\prime\prime}$ southwest of the nucleus,
respectively).  For simplicity, we refer to the latter three regions as
the ``nucleus,'' ``B1,'' and ``B2.''

For each region, Table~\ref{tab-Obs} lists the main constraints derived
in paper~I.  These include the intrinsic $K$-band and Lyman continuum
luminosities (\lk\ and \llyc)
\footnote{We use the following definitions: \lk\ is the luminosity
in the $K$ bandpass $\rm \lambda = 1.9 - 2.5~\mu m$ \citep{Wam81}
expressed in units of the total solar luminosity 
$\rm L_{\odot} = 3.85 \times 10^{26}~W$ while \llyc\ is the
Lyman continuum photon emission rate times an average ionizing
photon energy of 15~eV.},
the total bolometric luminosities from OB and cool evolved stars
as well as the separate contribution of OB stars (\lbol\ and \lbolOB),
the stellar masses (\mstars), the supernova rates (\snrate), the CO~1.62
and 2.29\,\micron\ bandhead equivalent widths (\ewcoh\ and \ewcoa),
and the neon fine-structure line ratio \neonteff.  The values of the
\lk/\llyc, \lbol/\llyc, \lbolOB/\llyc, \mstars/\lk, and \snrate/\lbol\
are also given.  Except for the starburst core, the neon ratio is an
``equivalent'' ratio as discussed below.

The $K$-band continuum luminosities \lk\ and the equivalent widths \ewcoa\
are corrected for the contribution and dilution by hot dust emission (which
is negligible for \ewcoh).  We assumed that 70\% of the OB stars' bolometric
luminosity is detected at IR wavelengths ($\rm \lambda = 5 - 300~\mu m$)
as reprocessed emission by dust, and that 30\% escapes perpendicular to
the galactic disk.  The stellar masses \mstars\ for the nucleus and the 
starburst core were computed by subtracting the mass of the gas component
(mainly in the form of $\rm H_{2}$) from the total dynamical mass.
Estimates of supernova explosion rates \snrate\ suffer from rather large
uncertainties (see paper~I), so related constraints will be used as consistency
arguments.  Among the nebular emission line ratios sensitive to the ionizing
OB stars available from our data sets, we chose \neonteff\ for this work.
For the starburst core, the ratio is that measured with the SWS, corrected for
extinction.  For the other regions, we applied the single-star photoionization
models of paper~I to derive ``equivalent'' neon ratios, which are those
predicted for the OB stars effective temperatures inferred from the
observed near-IR H to He recombination line ratios.

We also modeled individual regions corresponding to rebinned
$1^{\prime\prime} \times 1^{\prime\prime}$ pixels covering the 3D field of
view in its entirety.  The set of constraints for each pixel, from the 3D
data and the 12.4\,\micron\ map of \citet{Tel92}, includes the intrinsic \lk,
\llyc, \lbolOB, \ewcoa, and the equivalent neon ratio.  These constraints
are the most useful and relevant ones for the spatially-detailed modeling
based on the results of the selected regions.

\section{STARBURST MODELS} \label{Sect-models}

Our starburst models combine evolutionary synthesis of stellar clusters with
photoionization modeling of the surrounding gas.  They are optimized for
applications to $\rm \lambda = 1 - 45~\mu m$ observations of starburst
galaxies.  They assume that a given set of constraints applies to a
``homogeneous'' stellar population, i.e. either a single cluster
or an ensemble of clusters with identical IMF shape and cutoffs
\citep[see][]{Tho00}.

\subsection{Prediction of Stellar Properties} \label{Sub-STARS}

We computed the integrated properties of stellar clusters using the 
evolutionary synthesis code STARS \citep{Ste98, Tho00}.  STARS is
similar to other codes which have been presented in the literature
\citep[e.g.,][]{BC93, Cer94, Fio97, Lei99}.
STARS performs conventional synthesis, assigning fixed evolutionary
tracks to an appropriate set of mass bins.
STARS employs the Geneva stellar tracks \citep{Sch92}; for our analysis,
we selected the solar-metallicity tracks with normal mass-loss rates.
Although the tracks for enhanced mass-loss rates may provide better fits
to observations \citep[see][and references therein]{Mey94}, the choice has
little impact for this work.  The differences in the synthetized properties
of interest here are much smaller than those produced, e.g., by varying the 
IMF parameters and would imply changes $\rm \leq 1~Myr$ in the derived ages,
comparable to those from the measurement uncertainties.
We adopted a time-independent power-law IMF
${\rm d}N/{\rm d}m \propto m^{-\alpha}$ between lower and upper mass
limits \mlow\ and \mup.  The stars are assumed to form at a rate
$R(t_{\rm b}) = R_{0}\,e^{-t_{\rm b}/t_{\rm sc}}$, where $R_{0}$ is the initial
star formation rate (expressed in $\rm M_{\odot}\,yr^{-1}$), $t_{\rm b}$ is
the burst age, and $t_{\rm sc}$ is the burst decay timescale.

The integrated spectra of starburst galaxies in the $\rm 1 - 45~\mu m$ range
are dominated at the short-wavelength end by the direct light of cool evolved
stars and at the long-wavelength end by radiation of mostly OB stars that has
been reprocessed by interstellar dust and gas.  These populations are treated
with particular care in STARS.  Specifically, the non local thermodynamical
equilibrium (non-LTE) model atmospheres of \citet{Pau98}, which supersede
the earlier non-LTE atmospheres of \citet{Sel96}, are chosen to represent
adequately the spectral energy distribution (SED) of the hottest stars
\citep[see also][and references therein]{Pau01}.
Empirical photometric data are used to better account for the properties of
the coolest stars.  Finally, the Geneva tracks for intermediate-mass stars
have been extended to include the thermally-pulsing asymptotic giant branch
phase (TP-AGB), which can contribute significantly to the integrated $K$-band
luminosity.  STARS also predicts the strength of various near-IR stellar
absorption features.  Further details on the above aspects are given in
appendix~\ref{App-STARS}.

The \citet{Pau98} models include the non-LTE radiative transfer and
hydrodynamical treatment of radiatively-driven, spherically expanding
steady-state winds and account for the combined effects of line blocking
and blanketing on the radiative transfer and energy balance.
They provide excellent matches to observed high-resolution far-ultraviolet
spectra of hot stars \citep{Pau01} and nebular photoionization computations
which incorporate these model atmospheres successfully reproduce the relative
intensities of infrared emission lines measured in Galactic \ion{H}{2} regions
\citep{Giv02}.  As discussed by \citet{Giv02} and \citet{Tho00},
the inclusion of the \citeauthor{Pau98} models is critical
for the photoionization modeling because their SEDs are significantly
harder than for hydrostatic plane-parallel LTE models such as those of
\citet{Kur92}.
Accounting for the TP-AGB phase is most important for the CO bandheads
because it results notably in larger equivalent widths (EWs) at ages
$\rm \gtrsim 50~Myr$ \citep[see, e.g.,][]{Ori00}.  However, as will be
seen below, our results do not depend much on this feature.

\subsection{Prediction of Nebular Properties}  \label{Sub-CLOUDY}

We modeled the neon ratio produced in nebulae excited by the integrated
stellar SED computed by STARS using the photoionization code CLOUDY version
C90.05 \citep{Ferl96}.  The nebulae were represented as a single gas shell
surrounding a central point-like ionizing source.  In such ``central cluster''
models, the nebular conditions are specified by the gas and dust composition,
the hydrogen gas density $n_{\rm H}$, and the ionization parameter
$U \equiv Q_{\rm Lyc}/\left(4\pi r^{2} n_{\rm H} c\right)$, where
$Q_{\rm Lyc}$ is the intrinsic Lyman continuum photon emission rate
of the source, $r$ is the shell radius, and $c$ is the speed of light.  

We adopted the parameter values derived from the nebular analysis of paper~I:
$n_{\rm H} = 300~{\rm cm}^{-3}$, an {\em effective\/} ionization parameter
with $\log U_{\rm eff} = -2.3~{\rm dex}$, and solar photospheric gas-phase
abundances \citep[from][as implemented in CLOUDY v. C90.05]{Gre89, Gre93}.
As argued in paper~I, a more realistic representation of the \ion{H}{2}
regions in \mqd\ consists of a well-mixed distribution of ionizing clusters
and gas clouds.  The average ionization parameter at the outer boundaries
of the gas clouds derived from the observed properties assuming such a 
``random distribution'' model constitutes the appropriate effective value
for modeling in the framework of the idealized central cluster geometry.
Interestingly, we found nearly identical $\log U_{\rm eff}$ throughout
the starburst core of \mqd\ on scales of a few tens of parsecs to 
$\rm \sim 500~pc$.  As discussed in paper~I, this appears to indicate
a uniform star formation efficiency across the observed regions.  

Following \citet{Tho00}, we computed the
neon ratio for bursts with decay timescale $t_{\rm sc} = 1~{\rm Myr}$ by
assuming that $U_{\rm eff}$ varies proportionally with $Q_{\rm Lyc}$
and that the maximum value of $\log U_{\rm eff}$ equals $\rm -2.3~dex$.
For bursts with longer decay timescales, we convolved the corresponding
star formation rates $R(t_{\rm b})$ with the time-dependent neon ratio
for the short decay bursts.  We neglected the effects of dust grains
mixed with the ionized gas within the nebulae; this has little impact
on our model results (see \S~\ref{Sub-errors}).

\subsection{Modeling Procedure}  \label{Sub-proc}

We approached the modeling of \mqd\ by first making physical and qualitative
arguments to constrain the possible ranges for the parameters, and then
performed quantitative fits.  This led to physically plausible and successful
models although they may not be unique.
We first investigated, from the properties of selected regions, the general
characteristics of starburst activity: the IMF cutoffs and the typical burst
decay timescale (\S~\ref{Sect-models_regions}).  Having determined these
parameters and identified the most useful age indicators, we then modeled
individual regions across the entire 3D field of view to constrain the spatial
and temporal evolution of starburst activity (\S~\ref{Sect-models_pixels}).

Most of our constraints are sensitive to \mup, \tsc, and \tb,
so we investigated these parameters simultaneously.  The total stellar mass
depends only on the low-mass IMF (shape and \mlow), which we examined
independently.  We used ratios of different properties derived from data
obtained at similar wavelengths along with the CO bandheads EWs, minimizing
uncertainties from the extinction corrections and from the distance assumed
for \mqd.  The exception is \snrlbol\ but we used it mainly as consistency
check because of the uncertainties on \snrate\ (\S~\ref{Sect-Obs}).
We determined the initial star formation rates $R_{0}$ by comparing
the predicted luminosities to the measured values.  We kept the other
parameters constant; the effects of variations in the most critical ones
will be discussed when appropriate.  Table~\ref{tab-param} summarizes the
ranges considered or the fixed values adopted for the model parameters.

\section{MODELING OF SELECTED REGIONS IN M\,82}  \label{Sect-models_regions}

Figure~\ref{fig-mod_tsctb} compares the properties of the selected
regions from Table~\ref{tab-Obs} to model predictions.  The models were
computed for $m_{\rm up} = 100~{\rm M_{\odot}}$ and are shown for burst
decay timescales of 1~Myr, 5~Myr, 20~Myr, and 1~Gyr.  The diagnostics 
considered are not affected by the low-mass star population so a fiducial
$m_{\rm low} = 1~{\rm M_{\odot}}$ was here adopted.  This comparison reveals
age differences between the regions, in particular between the nucleus
(older) and the regions B1 and B2 (younger).  The exact ages depend on the
upper mass cutoff and burst decay timescale, which we constrained first.

\subsection{Upper Mass Cutoff and Burst Timescale}  \label{Sub-muptsc}

The neon and \lbol/\llyc\ ratios are our most sensitive diagnostics to
the upper mass cutoff.  Figure~\ref{fig-mod_mup} shows model computations
for these ratios with $m_{\rm up} = 25$, 30, 35, 50, and 100\masssol,
and for $t_{\rm sc} = 1$ and  5~Myr.
The upper mass cutoff is best constrained from the properties of the young
massive stars only.  Therefore, the observed bolometric luminosities for
selected regions in Figure~\ref{fig-mod_mup} exclude the contribution from
cool evolved stars so that the \lbolOB/\llyc\ ratios are compared to the
models.

Both diagnostics imply similar dominant OB star populations for the various
regions considered.  The small values of the neon ratio suggest a relative
paucity of ionizing photons with energies $\rm > 3~Ryd$ (the ionization
potential of $\rm Ne^{+}$) while the low \lbolOB/\llyc\ ratios imply that
lower energy Lyman continuum photons are being produced efficiently.  From
the comparison with the models, two alternative interpretations are possible:
(1) a high $m_{\rm up} \gtrsim 50~{\rm M_{\odot}}$, a short timescale of at
most a few million years, and the softening of the ionizing radiation field
attributable to rapid aging of the starburst, or
(2) a lower \mup\ down to $\approx 30~{\rm M_{\odot}}$, with longer
timescales possible.  
Upper mass cutoffs below 30\masssol\ are ruled out, but the data do not
allow us to constrain unequivocally $m_{\rm up}$ and $t_{\rm sc}$ for
$m_{\rm up} \geq 30~{\rm M_{\odot}}$.

As summarized by \citet{Tho00}, a variety of results on local
templates of starburst regions as well as other large-scale starburst
systems make high upper mass cutoffs more plausible.  In \mqd, the
low \lbolOB/\llyc\ values for B1 and B2 can only be reproduced with
$m_{\rm up} \gtrsim 50~{\rm M_{\odot}}$.  The neon and \lbolOB/\llyc\ ratios
measured for all selected regions lie within the ranges determined for the
sample of dusty solar-metallicity IR-luminous starburst galaxies of
\citet{Tho00}.   These ranges, indicated in Figure~\ref{fig-mod_mup}, are
compatible with $m_{\rm up} \geq 50~{\rm M_{\odot}}$ and differences among
the sample sources attributable to a range in burst ages. 

In addition, the poor correlation in \mqd\ on scales of a few tens of
parsecs between the spatial distributions of the near-IR continuum
emission and CO bandhead EWs, and that of the ionized gas
(e.g., \brg\ and \ion{He}{1} 2.06\,\micron) and IR dust emission provides
strong evidence for short decay timescales in individual burst sites
(see the 3D images of paper~I; also, e.g., \citealt{Tel91, Lar94, Sat97}).
Since red supergiants dominate the near-IR continuum emission in \mqd\
(\S~\ref{Sub-age}), the poor correlation implies that star formation activity
was not maintained locally at high levels longer than the typical main-sequence
lifetime of the most massive red supergiants progenitors ($\rm \approx 5~Myr$).
This situation is reminiscent of the nucleus of M\,83, where \citet{Pux97}
found that the K-band continuum emission and CO bandhead strength are
spatially anticorrelated with the \brg\ EW, which they interpreted, together
with starburst models, as indicating an evolutionary sequence and burst
timescales of $\rm 1 - 5~Myr$.  Massive star formation is expected to inhibit
further star formation locally already after a few million years owing to
strong stellar winds and supernova explosions disrupting the surrounding
interstellar medium (ISM).  The $\rm \sim 100~pc$--size gas hole around the
brightest radio supernova remnant $41.9+58$ seems to provide a particularly
striking example of such effects in \mqd\ \citep[e.g.,][]{Wei99}.

The most remarkable outcome of our models is that for
$m_{\rm up} \geq 50~{\rm M_{\odot}}$, the OB star-dominated properties of
the 3D field of view and the starburst core imply short burst decay timescales
on large spatial scales ($\rm \sim 100 - 500~pc$).  This is not due to a few
regions dominating the integrated properties; for instance, B1 and B2
contribute together less than 15\% to the \llyc\ and \lbolOB\ measured in the
3D field of view.  Our results suggest that in \mqd\ starburst activity can
have strong negative feedback effects globally as well.  We discuss this
point further in \S~\ref{Sect-discussion}.

\subsection{Successive Starburst Events}  \label{Sub-2bursts}

In the rest of this paper, we will adopt $m_{\rm up} = 100~{\rm M_{\odot}}$
and a representative $t_{\rm sc} = 1~{\rm Myr}$ as justified above.
Examination of Figures~\ref{fig-mod_tsctb} and \ref{fig-mod_mup} then reveals
that for each region, there is a systematic increase in burst age from that
implied by the tracers of OB stars (neon and \lbolOB/\llyc\ ratios) to that
inferred from the tracers of cool evolved stars (CO bandheads).  
The differences are about 5~Myr.
The \lk/\llyc\ ratios, measuring the relative populations of these stellar
components, correspond to intermediate ages.  

Figure~\ref{fig-modreg} illustrates the difficulty
of single burst models to meet simultaneously the constraints for each of the 
selected regions.  The figure compares the regions' properties to evolutionary
tracks for single bursts in diagrams of the neon, \lbol/\llyc, and \lk/\llyc\
ratios versus \ewcoa.  At ages corresponding to the measured values of \ewcoa,
models for $t_{\rm sc} = 1~{\rm Myr}$ and $m_{\rm up} = 100~{\rm M_{\odot}}$
predict neon ratios 10 (B1) to 34 (nucleus) times lower than measured,
\lbol/\llyc\ ratios 6 (nucleus) to 23 (B2) times higher, and \lk/\llyc\ ratios
4 (nucleus) to 33 (B2) times higher.  The constraints are still not satisfied
simultaneously when augmenting the burst decay timescale to 5~Myr with,
in particular, predicted neon ratios about an order of magnitude higher
than measured at ages corresponding to the observed values of \ewcoa.
Allowing for lower $m_{\rm up}$ down to 50\masssol\ only worsens the
discrepancies.  

The age sequence between the various diagnostics may suggest that the hot
massive stars and the cool evolved stars belong to populations formed in
distinct, successive starburst events.  In this scenario, the ``young bursts''
account for most of the ionizing radiation while the ``old bursts'' dominate
the near-IR properties.  \citet{Rie93} first proposed such two-burst
models of the entire starburst core of \mqd\ for similar reasons.
We thus considered models consisting of two sequential bursts, each
with $t_{\rm sc} = 1~{\rm Myr}$ and $m_{\rm up} = 100~{\rm M_{\odot}}$,
and separated in time by more than one decay timescale.  
Figure~\ref{fig-modregtwo} shows similar evolutionary tracks as 
Figure~\ref{fig-modreg} for the best-fit time separation and relative
burst intensities for each region derived in \S~\ref{Sub-age}.
From Figures~\ref{fig-modreg} and \ref{fig-modregtwo}, it is also
difficult to disentangle between single and two-burst models.  However,
since the neon and \lbolOB/\llyc\ ratios seem to favour the shortest decay
timescales, successive short bursts could provide a natural explanation
for the observed strength of the CO bandheads.  The two-burst scenario also
appears plausible in view of the morphology and nearly edge-on orientation
of \mqd\ (\S~\ref{Sect-discussion}).

We remark that given the uncertainties of the data and models, it may be
possible to find single bursts with finely-tuned parameters which reproduce
all the observed properties.  Exponentially decaying bursts with \tsc\
intermediate between 1 and 5~Myr, for instance, could meet the neon ratio
constraint while violating the \lbol/\llyc\ and \lk/\llyc\ constraints
at a level that is more severe than for the double bursts but probably
still acceptable.  As another example, allowing lower \mup\ down to
30\masssol\ together with longer \tsc\ of $\rm 10 - 20~Myr$ for single
bursts may improve the fit to the neon ratios but the mismatch is only
redistributed to the other properties.  Similar difficulties in modeling
comparable sets of constraints have been encoutered for other dusty
IR-selected starbursts, e.g., by \citet{Eng98} for NGC\,253 and
\citet{Gol97} for a sample of (ultra)luminous IR galaxies.
With our data and models of \mqd, together with the obvious complexity
of the problem, it is not possible to find a unique solution.

\subsection{Burst Age and Strength}  \label{Sub-age}

Adopting the two-burst scenario as working hypothesis, we constrained the
age of the young bursts from the neon and \lbolOB/\llyc\ ratios, and of the
old bursts from the CO bandheads.  The initial star formation rates $R_{0}$
(with \mlow\ fixed at 1\masssol) were adjusted so that the total contribution
from both bursts to \llyc\ and \lk\ reproduces the observed values.  For
identical IMF parameters and burst decay timescales, the values of $R_{0}$
provide a meaningful measure of the relative strengths of the bursts.

For the young bursts, the ages inferred from each indicator agree marginally,
the neon ratio implying slightly older ages.  This discrepancy would not be
solved by other choices of \mup\ and \tsc.  The difference is comparable to the
burst timescale and, in view of the measurements and modeling uncertainties,
is probably not significant.  
The \lbolOB/\llyc\ ratio may trace a slightly different population as it
is sensitive to a somewhat lower stellar mass range than the neon ratio is.
The effects of a distribution of cluster masses and luminosities or of dust
grains within the nebulae, neglected here, would increase the discrepancy
(\S~\ref{Sub-errors}).
We therefore assigned equal weights to the diagnostic ratios.

For the old bursts, an accurate age determination is only possible for the
nucleus.  It has very deep CO bandheads which the models reproduce only at ages
of $8 - 15~{\rm Myr}$, consistent with the average spectral type \ion{K5}{1}
derived from detailed spectral synthesis in paper~I.  The EWs at the nucleus
further support short burst timescales: for $t_{\rm sc} \geq 5~{\rm Myr}$,
the contribution from older populations with shallower bandheads and/or
photospheric and nebular featureless continuum emission from OB stars
results in predicted EWs lower than observed, for any burst age.  
Figure~\ref{fig-EWs} indicates that older ages when intermediate-mass
stars reach the end of the asymptotic giant branch
($\sim 50~{\rm Myr} - 1~{\rm Gyr}$) are also ruled out from \ewcoa.

For the other selected regions, we can set firm lower limits on
the ages but the EWs alone do not provide strong constraints on
upper limits.  For comparison with our data and models, we indicate
in Figure~\ref{fig-EWs} the ranges observed by \citet{OOKM95} for 
old ($1 - 10~{\rm Gyr}$) stellar populations characteristic of the
central regions of elliptical galaxies and bulges of spiral galaxies
(hereafter ``normal populations'').  The EWs at B1 are somewhat lower
than for normal populations and may indicate an early stage when
the first red supergiants appear.  For B2, the 3D field of view, and
the starburst core, they are slightly higher and imply possible upper
limits on the age of $\approx 50~{\rm Myr}$ and on the timescale of
$\approx 5~{\rm Myr}$.

The \mlk\ ratio provides an additional constraint.  Empirical
determinations for normal populations lie in the range 
$10 - 30~{\rm M_{\odot}/L_{\odot}}$ \citep*{Dev87, OOKM95, Hun99}.
\footnote{Our units for the \mlk\ ratio differ from the conventional
definition expressed in $(M/L_{K})_{\odot}$.  The quoted range has been
computed in $\rm M_{\odot}/L_{\odot}$ from the data presented in the 
given references, assuming a $K$ bandpass $\rm \lambda = 1.9 - 2.5~\mu m$
and expressing \lk\ in $\rm L_{\odot} = 3.85 \times 10^{26}~W$.}
As argued in paper~I, the substantially lower values 
$\sim 1~{\rm M_{\odot}/L_{\odot}}$ measured for \mqd\ between radii of
$\rm \approx 10$ and 250~pc indicate very large contributions from red
supergiants to the near-IR continuum emission, even in regions of
low surface brightness.  Consequently, ages $\lesssim 50~{\rm Myr}$ are
the most plausible.  We adopted the solutions corresponding to the rising
part of the predicted EWs with $t_{\rm sc} = 1~{\rm Myr}$ but older ages
of $\rm 15 - 30~Myr$ are also possible.  Our choice implies lower $R_{0}$
for the old bursts since the predicted \lk\ peaks around 10~Myr.
The younger age solutions are unaffected by uncertainties related to
numerical fluctuations due to our conventional synthesis technique and
to our TP-AGB implementation, which may become a concern at ages
$\rm \gtrsim 50~Myr$ (see appendix~\ref{App-STARS}).

\subsection{Model Results for Selected Regions}  \label{Sub-resreg}

Tables~\ref{tab-modreg} and \ref{tab-modglob} report our two-burst model
results for the selected regions.
The derived burst strengths outline the particular intensity of starburst
activity in the central few tens of parsecs of \mqd\ about 10~Myr ago.
Our model for the starburst core agrees well with the results of \citet{Rie93}
who found that separate bursts occurring about 5 and 13~Myr ago, each lasting
a few million years, could reproduce the observed properties.  The small
differences between our model and theirs reflect mainly differences in the
adopted set of constraints and model assumptions (such as their choice of
Gaussians to represent the star formation rate).

We estimated the formal uncertainties on the relative burst ages and
strengths $R_{0}$ by varying in turn the observational constraints within
the ranges corresponding to the measurement errors.  They are typically 
$\rm \lesssim 0.5~Myr$ and $25\% - 40\%$ for the ages and strengths, 
respectively.  For the old bursts, the minimum age is fairly well constrained
by the CO bandhead EWs whereas the maximum age from the quoted uncertainties
constitutes a lower limit since we did not account here for the older possible
solutions.  The absolute burst ages and intensities may suffer from larger
uncertainties related, e.g., to the burst decay timescale and shape of the
star formation rate.  The relative ages and intensities are less affected
because the evolution of the properties considered are dominated by stellar
evolution during the early phases of a burst and for decay timescales
comparable to the main-sequence lifetime of the most massive stars.

Figure~\ref{fig-resreg} illustrates the results for the most relevant
properties: \lk, \llyc, \lbol, \ewcoa, \ewcoh, and the neon ratio.  For
each region, the curves show the variations of the properties as a function
of the time elapsed since the onset of the old burst and are normalized
to the observed values.  They represent the combination of the two bursts,
weighted by the respective values of $R_{0}$ for each burst.  The models for
all regions reproduce the main constraints within a factor of three or better.
The masses computed for the
nucleus and the starburst core are, however, substantially lower than the
measured values.  The mass is very sensitive to the lower cutoff and shape
of the IMF, which we discuss in \S~\ref{Sub-mlow}.

Our results are roughly consistent with the constraints
imposed by \snrate\ and $10^{12}\,\nu_{\rm SN}/L_{\rm bol}$.  
A good agreement for the regions observed with 3D is not expected
because of the large uncertainties in deriving \snrate\ from
[\ion{Fe}{2}] 1.644\,\micron\ line fluxes, as discussed in paper~I.
The ages inferred for the old bursts satisfy the ``timing constraint''
implied by the extent of the X-ray halo tracing the supernova-driven
wind of \mqd, which requires a minimum starburst age of about 10~Myr
\citep*{Hec90, Rie93, Leh99}.

\subsection{Model Uncertainties}  \label{Sub-errors}

We investigated the effects of various sources of uncertainties on our model
results so far, including the assumption of a homogeneous cluster population,
the nebular parameters, the shape of the IMF, and the synthetic model
atmospheres.   Their effects have been described for the neon ratio
by \citet{Tho00} whom we refer to for further details.
Figure~\ref{fig-paramvar} shows in addition the effects on the \lbol/\llyc\
and \lk/\llyc\ ratios, and on \ewcoa. 
None of the above sources of uncertainties affect our conclusions concerning
the upper mass cutoff of the IMF or the ages and relative intensities derived
for the bursts.  While possible alternative assumptions or variations of the
nebular parameters and IMF shape result in measurable differences in the
model predictions, the effects remain small compared to those of aging
or changes in \mup.

In particular, the largest effects are obtained for a
``heterogeneous'' population of stellar clusters following the
luminosity function (LF) derived by \citet{Tho00}
\footnote{For reference, the LF in terms of
cluster H ionizing rate $Q^{\rm cl}_{\rm Lyc}$ is
${\rm d}N/{\rm d}\left(\log Q^{\rm cl}_{\rm Lyc}\right) \propto 
\left(Q^{\rm cl}_{\rm Lyc}\right)^{-\beta}$, with $\beta = 0.19$ for
$Q^{\rm cl}_{\rm Lyc} = 10^{45} - 10^{49.5}~{\rm s^{-1}}$ and $\beta = 1$
for $Q^{\rm cl}_{\rm Lyc} = 10^{49.5} - 10^{53}~{\rm s^{-1}}$.}.
This LF was constructed assuming it reflects a cluster mass function and
imposes a distribution in effective \mup\ of clusters (at fixed IMF slope
and \mlow).  The large proportion of small clusters results in a
``down-weighting'' effect with, for instance, lower neon ratios and shallower
CO bandheads due to the reduction in relative population of massive stars.
However, the effects are but a factor of two or less.
Similar differences for the neon ratio are obtained by
varying the ionization parameter within the possible range 
$-2.5 \leq \log U_{\rm eff} \leq -2$ for \mqd\ (paper~1)
or adopting the softer \citet{Kur92} SEDs for hot stars instead
of the \citet{Pau98} SEDs \citep[see][]{Tho00}.
The effects of dust within the \ion{H}{2} regions, illustrated
with the Orion nebular composition, are significantly smaller.
By absorbing ionizing photons, dust would also reduce the \llyc\
in the models.  This is difficult to quantify, but the resulting
increase in synthetized \lbol/\llyc\ and \lk/\llyc\ would only imply
younger ages and keep the inferred burst decay timescales short.

Figure~\ref{fig-IMF} shows that the alternative IMFs we considered,
from \citet{MS79} and that derived by \citet{Eis98} for the
Galactic high-mass star-forming region NGC\,3603, bracket the range
of slopes at intermediate and high masses determined for young clusters
and OB associations in the Milky Way and Large Magellanic Cloud
(e.g., \citealt{Hun97} and references therein; 
\citealt{Bra96, Mas98, Kro01}).  Changes in the IMF shape have
the smallest effects in our models.   As a consequence, our set of 
observational constraints does not allow us to investigate the shape
of the intermediate- and high-mass IMF in \mqd.

Potential problems with the stellar evolutionary tracks represent an
additional source of uncertainties.  A detailed discussion is beyond the
scope of this paper but we briefly mention two points relevant to this work.
Origlia and co-workers \citep{Ori99, Ori00} warned that model 
predictions of the CO bandheads can be significantly underestimated
at ages $\rm \gtrsim 10~Myr$ if the TP-AGB phase is not accounted for
\citep[see also][]{CB91, LRV94} and because of the possible inadequacy
of current stellar tracks for the evolution of massive stars in the red
supergiant phase, most severely at sub-solar metallicities.
Our starburst models account for the TP-AGB (see appendix~\ref{App-STARS})
and use solar-metallicity tracks as appropriate for \mqd\ (paper~I),
reducing the impact of these uncertainties on our results.
Furthermore, the maximum age from the CO bandheads would be most affected,
and much less the minimum age so that the above uncertainties have little
consequences on the age sequence suggested by our observational constraints.
This result would be more affected if the stellar tracks significantly
underpredict the number of very luminous but relatively cool stars
($\log L_{\rm bol} > 5.75$, $\log T_{\rm eff} < 4.5$), a possibility
raised by \citet{Tho00} in their discussion of the Galactic Center.  With
correspondingly lower predicted neon ratios, the constraints might be more
clearly accomodated with single bursts of somewhat longer durations.
However, this point remains to be thoroughly investigated.

\subsection{Low-mass End of the IMF}  \label{Sub-mlow}

The low-mass IMF for the starburst population in \mqd\ has long been debated.
Various authors argued that it is deficient in stars with initial masses below
$\approx 3~{\rm M_{\odot}}$ compared to the solar neighbourhood IMF
\citep{Rie80, Rie93, Ber92, Doa93}.  The studies
by Rieke and coworkers have been the most influential.  \citet{Rie80}
argued for a sharp truncation of the IMF at 3.5\masssol, with no stars formed
below this cutoff.  \citet{Rie93} modified this conclusion and argued that
while the IMF may extend down to 0.1\masssol, the slope flattens significantly
near 3\masssol, at a substantially higher mass than in the solar neighbourhood
(near 0.5\masssol; see, e.g., \citealt{Kro01} and references therein).
More recently, \citet{Sat97} challenged these hypotheses and modeled the
integrated $K$-band luminosity of \mqd\ with a Salpeter IMF down to
0.1\masssol\ without using up more than $\approx 35\%$ of the total 
dynamical mass in the starburst, a reasonable upper limit.

The origin of the controversy resides primarily in the notable differences
in observational constraints and model assumptions between the various
studies.  Here, we reexamine the issue of the low-mass IMF in \mqd\ in light
of our new data and models.  We consider two regions for this purpose: the
nucleus (central 35~pc) and the starburst core (central 500~pc).

\subsubsection{Reexamination of the Low-mass IMF}

Among our constraints, the stellar mass \mstars\ is the only one 
sensitive to the low-mass star population.  We computed this critical
quantity by subtracting the mass of the gaseous component $M_{\rm gas}$
from the total dynamical mass $M_{\rm dyn}$ (see paper~I for details).
We derived the enclosed $M_{\rm dyn}$ as a function of projected radius
from position-velocity maps of the millimeter CO $J = 1 \rightarrow 0$,
[\ion{Ne}{2}] 12.8\,\micron\, and [\ion{S}{3}] 9069\AA\ line emission
\citep{She95, Ach95, McK93} combined with the result of stellar velocity
dispersion measurements at the nucleus based on the CO 2.29\,\micron\ bandhead
by \citet*{Gaf93}.  We estimated $M_{\rm gas}$, accounting for molecular
and ionized gas, from the CO $J = 1 \rightarrow 0$ map of \citet{She95}
using the CO intensity to $\rm H_{2}$ column density conversion factor
derived by \citet{Wil92} from multi-line radiative transfer analysis and
from the Lyman continuum luminosities.
The \mstars\ for the nucleus and the starburst core are $7.9 \times 10^{7}$
and $\rm 6.1 \times 10^{8}~M_{\odot}$, respectively.

We investigated the low-mass IMF (shape and cutoff) by comparing the mass
of stars predicted by the model bursts \massstb\ with the \mstars\ derived
for each region.
In this comparison, allowance has to be made for a possible stellar population
predating the starburst, which can contribute to the measured stellar mass, so
that the computed \massstb\ does not exceed a plausible fraction of \mstars.
To estimate this fraction, we followed the argument of \citet{McL93}.
Tidal interaction with M\,81 presumably induced the starburst in \mqd;
numerical simulations indicate that the perturbations resulting from such
encounters can efficiently drive large amounts of gas towards the nuclear
regions and that before the gas mass equals the mass in preexisting stars,
self-gravitation in the gas should trigger a starburst with short timescale
compared to the gas settling time \citep[e.g.,][]{Her89, Mih94, Mih96}.
Thus, \massstb\ unlikely exceeds half of $M_{\rm dyn}$ in \mqd.  In our
analysis, we adopted a conservative upper limit for \massstb\ of 50\% of
the total {\em stellar\/} mass \mstars.  We did not account for the gas mass
returned to the ISM via stellar winds and supernova explosions, or expelled
out of the galaxy in the starburst wind.  Since massive stars contribute most
to the mass return for the burst ages $\rm \lesssim 20~Myr$ relevant here but
do not dominate \massstb, these effects can be neglected to first order but
will be further addressed below.

Our two-burst models for the nucleus and the starburst core
(Tables~\ref{tab-modreg} and \ref{tab-modglob}) use only 9\% and 33\%, 
respectively, of the stellar masses for a Salpeter IMF slope between
$m_{\rm low} = 1~{\rm M_{\odot}}$ and $m_{\rm up} = 100~{\rm M_{\odot}}$,
but this choice of \mlow\ was arbitrary.  Figure~\ref{fig-mlow} shows the
burst masses required to reproduce the observed luminosities (\lk\ and \llyc)
computed for \mlow\ in the range $\rm 0.1 - 5~M_{\odot}$, with a
\citeauthor{Sal55} slope and for the same burst ages and decay timescales.
For the nucleus, the bursts never use up more than 30\% of the
measured \mstars.  On the other hand, the burst mass for the starburst core
reaches 50\% of \mstars\ at $m_{\rm low} = 0.4~{\rm M_{\odot}}$ and equals
\mstars\ at $m_{\rm low} \approx 0.1~{\rm M_{\odot}}$.  For the mass involved
in the starburst not to exceed our adopted limit would require a sharp cutoff
at 0.4\masssol\ for a \citeauthor{Sal55} IMF.  Alternatively, allowing the IMF
to extend down to 0.1\masssol, a flattening of the IMF at low masses would be
required to reduce the increase in \massstb\ with decreasing \mlow\ so that
$M^{\star}_{\rm stb} \leq 50\%$ of $M^{\star}$.  
With only one constraint available, our analysis does not allow us to break
the degeneracy between the cutoff and the shape of the low-mass IMF.

\subsubsection{Factors Influencing the Low-mass IMF Determination}

The \massstb\ predicted by the models is determined by the absolute
luminosities.  This makes inferences on the low-mass IMF very sensitive to
the extinction corrections applied, to the star formation history, and to
the shape of the IMF at higher masses.
The first point has been extensively discussed by \citet{McL93}, \citet{Rie93},
and \citet{Sat97}.  In that respect, our intrinsic luminosities correspond
to the higher ranges reported in the literature and therefore impose
more stringent constraints.  In particular, the \lk\ used by \citet{Sat97}
and the lower limit considered by \citet{Rie93} are both smaller than our
value ($5 \times 10^{8}$ and $\rm 7.8 \times 10^{8}~{\rm L_{\odot}}$
compared to $1.3 \times 10^{9}~{\rm L_{\odot}}$ for a 
$\lambda = 1.9 - 2.5~{\rm \mu m}$ $K$ bandpass and the photometric system
of \citealt{Wam81} we adopted).
With our two-burst models, they allow \mlow\ down to 0.2\masssol\ or
less, as shown in the bottom panel of Figure~\ref{fig-mlow} (the mass
constraints are similar between all three studies).

The star formation history can also significantly alter the conclusions
about the low-mass IMF since longer or older bursts tend to use up more mass.
In this regard, however, our conclusions above for a Salpeter IMF differ from
those of \citet{Sat97} mainly because of the different \lk\ adopted.
In terms of \massstb, their single burst model for the starburst core, with
an age of 10~Myr and a short duration, is comparable to our two-burst model
in which the old burst with an age of 9~Myr contributes 75\% of the total
burst mass for this region.  Comparison with the models of \citet{Rie93} is
less straightforward; our two-burst model is very similar to theirs with
respect to burst ages and durations but the adopted IMFs are very different.

For the young ages of interest here, the various luminosities are dominated
by high-mass stars, so that the relative slope of the IMF between different
mass ranges is particularly critical in constraining the low-mass IMF.  For
instance, IMFs which steepen at high masses are less efficient at converting
the mass into luminosity.  In order to reproduce the luminosities, they
require larger \massstb. 
To illustrate this, we considered three alternative IMFs:
the \citet{MS79} IMF for the solar neighbourhood, the IMF proposed by
\citet{Rie93} for \mqd\ (their ``IMF~8''), and the IMF of \citet{Eis98}
for NGC\,3603.  \citeauthor{Eis98} derived the NGC\,3603 IMF based on
near-IR adaptive optics observations down to 1\masssol, their detection limit.
For our purposes, we allowed extensions below 1\masssol\ using the slope
determined in the $\rm 1 - 15~{M_{\odot}}$ range.  
Table~\ref{tab-IMF} gives the parameters of the IMFs (see also
Figure~\ref{fig-IMF}) and Figure~\ref{fig-mlow} shows the variations
of the burst masses versus \mlow\ computed with each of them.

For the range $0.1 - 100~{\rm M_{\odot}}$, the mass fraction locked in stars
less massive than 1\masssol\ is significantly smaller for the \citeauthor{MS79}
IMF compared to the \citeauthor{Sal55} IMF (44\% and 61\%), while the
corresponding fractions for stars more massive than 10\masssol\ are nearly
identical (8\% and 12\%, respectively).  However, properties such as \lk, \lbol,
and \llyc\ are steeply increasing functions of the stellar mass and the
\citeauthor{MS79} IMF is much steeper at the high-mass end than the
\citeauthor{Sal55} IMF.  Consequently, normalizing to the same luminosities,
the \citeauthor{MS79} IMF predicts substantially more mass than the
\citeauthor{Sal55} IMF, with \massstb\ larger by up to a factor of two
in our models.   The \citet{Rie93} IMF is very similar to the \citeauthor{MS79}
IMF but with the first inflection point moved up from 1\masssol, giving
particular emphasis on the intermediate-mass range.
Of the various IMFs they explored, such modifications to solar-neighbourhood
type IMFs were the most successful at reproducing their set of constraints.
Finally, compared to the \citet{Rie93} IMF, that of \citet{Eis98} emphasizes
less the intermediate-mass range but has a larger proportion of stars above
10\masssol\ (40\% versus 23\% of the total mass between 1 and 100\masssol),
implying much higher mass-to-luminosity conversion efficiency.

Figure~\ref{fig-mlow} shows that for the central 35~pc of \mqd, the 
low-mass IMF remains unconstrained for any of the three alternative IMFs: 
the computed \massstb\ never exceeds 35\% of the measured \mstars\ with 
\mlow\ down to 0.1\masssol.  The case of the starburst core is again more
restrictive.  With the \citet{MS79} IMF, more than 50\% of the stellar mass
is involved in the bursts for $m_{\rm low} \leq 2~{\rm M_{\odot}}$.  
If we allow for stars down to 0.1\masssol\ to form, we conclude as
\citet{Rie93} that a displacement of the first inflection point to higher
masses is then needed to effectively reduce \massstb.  
With our data and models, though, \massstb\ still exceeds half of $M^{\star}$
at $m_{\rm low} \leq 1~{\rm M_{\odot}}$ if we adopt the \citeauthor{Rie93} IMF.
The \citet{Eis98} IMF allows us to
extend the IMF with the same $\alpha = 1.73$ power-law index from 15\masssol\
down to 0.1\masssol\ without using up more than 30\% of the stellar mass
in the bursts.  Interestingly, if we assume both the IMF and the $K$-band
luminosity of \citet{Rie93}, the \massstb\ remains well below 50\% of
$M_{\star}$ over the entire range of \mlow\ (dashed line in the bottom panel
of Figure~\ref{fig-mlow}).

Another consideration worth pointing out is the mass loss via massive stars
winds and supernova explosions, which is important already within a few
million years of the onset of star formation \citep[see, e.g.,][]{Lei99}.
In particular, the mass returned to the ISM could have been entrained out of
\mqd\ in the starburst wind, in which case our $M^{\star}$ constraint would
be missing some fraction or, alternatively, the predicted \massstb\ curves
of Figure~\ref{fig-mlow} would decrease accordingly. An estimate
of the effect is difficult because of the unknown amount of returned
mass actually driven away in the starburst wind.  As an illustration, we may
consider the extreme case in which massive stars $\rm \gtrsim 10~M_{\odot}$
lose all their mass within $\rm \approx 20~Myr$, and that all ejectae are
carried out of \mqd\ by the starburst wind.  For a Salpeter IMF extending
from 1 to 100\masssol, the range $\rm \geq 10~M_{\odot}$ accounts for 31\%
of the total mass so that at most $\approx 30\%$ of \massstb\ would be lost
and expelled.  This fraction varies with \mlow, up to $\approx 50\%$ at
3\masssol\ and down to $\approx 10\%$ at 0.1\masssol, and more generally
depends on all IMF parameters.  The overall impact is to relax the
requirements on the low-mass IMF but the effect is not likely to be
more important than the other factors discussed above.

\subsubsection{Additional Remarks}

In view of the uncertainties involved and the lack of strong constraints 
on the shape of the IMF, the low-mass IMF in \mqd\ remains elusive and only
general statements can be made.  If stars formed down to 0.1\masssol, our
data and models may indicate a substantial flattening of the starburst IMF
below a few $\rm M_{\odot}$, assuming a Salpeter slope at higher masses.
This requirement would be alleviated for a flatter intermediate- and high-mass
IMF or for less conservative limits on \massstb.  A bias against the formation
of low-mass stars has been suggested for other starburst galaxies as well
\citep*[e.g.,][]{Aug85, Wri88, Olo89, Nak89, Pre94}.
\citet{Eng96, Eng98} concluded for NGC\,6946 and NGC\,253 that the IMF
proposed by \citet{Rie93} for \mqd\ is more compatible with the observations
than a solar-neighbourhood IMF, while \citet{Alo01} found that the IMF in
NGC\,1614 may be even more biased towards high-mass stars.

Kinematic mass-to-light ratios have been obtained for a few individual
luminous stellar clusters in \mqd, from velocity dispersion measurements
combined with high resolution {\em Hubble Space Telescope\/} ({\em HST\/})
imaging.  Using optical data obtained for the super star cluster ``M\,82-F''
$\rm \approx 450~pc$ southwest of the nucleus and an age estimate of 60~Myr,
\citet{Smi01} concluded that the cluster IMF is top-heavy and requires a
low-mass cutoff at $\rm 2 - 3~M_{\odot}$ for an $\alpha = 2.3$ slope.
\citet*{MCr03} studied two younger ($\rm \sim 10^{7}~yr$) clusters
roughly 180~pc southwest of the nucleus based on near-IR observations.
For an IMF with $\alpha = 1.3$ at $\rm 0.1 - 0.5~M_{\odot}$
and $\alpha = 2.3$ at $\rm 0.5 - 100~M_{\odot}$
\citep[based on][]{Kro01}, their data support a relative deficiency
at $\rm \lesssim 1~M_{\odot}$ for one cluster but allow \mlow\ down to
0.1\masssol\ for the other.  In their revision of the \citet{Smi01} analysis 
of M\,82-F, using the same mass but combined with the near-IR luminosity
and a lower age of 40~Myr, they confirmed the evidence for a top-heavy IMF
although with a looser requirement of $m_{\rm low} \approx 1~{\rm M_{\odot}}$.
Similar investigations of young super star clusters in other systems, 
including NGC\,1569A and NGC\,1701-1 \citep{Ste98}, and five clusters
in NGC\,4038/4039 \citep{Men02}, yielded mixed results as well.
This seems to indicate a range of possible IMFs in super star clusters
which may be related to their birth environment or to mass segregation 
\citep[see, e.g.,][]{Men02, MCr03}.  The results of the \mqd\ clusters may
provide support in favour of an IMF flattening at low masses but, as pointed
out by \citet{Smi01}, it is unclear whether conclusions about luminous star
clusters, let alone of a handful only, can be extended to the overall pattern
of star formation in the starburst.

Direct censuses of the low-mass star populations in Galactic and
near-extragalactic high-mass star-forming regions are still scarce and
difficult owing to the faintness of such stars and to important crowding
effects.  Recent results from very high sensitivity and angular resolution
observations suggest that mere truncations of the IMF near 1\masssol\ are not
plausible, but the shape of the low-mass IMF remains poorly constrained.  In
particular, \citet{Bra99} found that the central parsec of NGC\,3603 is well
populated by pre-main sequence stars down to 0.1\masssol, based on data from
the Very Large Telescope.  The near-IR adaptive optics observations by
\citet{Eis98} show no evidence for a significant flattening of the IMF down
to their detection limit of 1\masssol.  In contrast, from deep {\em HST\/}
images of the R\,136 cluster in 30~Doradus, \citet{Sir98, Sir00} could probe
the $\rm 1.35 - 6.5~M_{\odot}$ range and reported a flattening of the IMF
below 2\masssol, a result however disputed by \citet{Sel99}.
Given the shape of the solar-neighbourhood IMF, with possible turnover near
0.5\masssol\ \citep[e.g.,][]{Sca86, Sca98, Ran87, Kro01}, and
indications of a flattening towards lower masses in some local templates
of starburst regions and super star clusters, our inferences for \mqd\ may
not need be interpreted in terms of an ``abnormal'' deficiency of the IMF
in low-mass stars.

\section{SPATIALLY DETAILED MODELING OF M\,82} \label{Sect-models_pixels}

The modeling of selected regions in the previous section provides
the general characteristics of starburst activity in the central
regions of \mqd, but only an incomplete picture of its evolution.
In particular, our choice of the small-scale regions as those with
brightest continuum or line emission, and with deepest or shallowest
CO bandheads likely introduced a bias towards preferential ages.  In
order to obtain a more complete picture, we modeled 
{\em individual regions throughout the entire 3D field of view}. 
These have a range in their properties suggesting a range in evolutionary
states and burst strengths.  Although along any line of sight the 
integrated properties will always be dominated by the most luminous
populations, a detailed modeling on small spatial scales reduces the
bias towards particular ages.

Given our results of \S~\ref{Sect-models_regions}, we assumed that a Salpeter
IMF with high upper mass limits and burst decay timescales $\rm \lesssim 5~Myr$
are appropriate for individual pixels.
We also assumed that the near-IR properties are dominated everywhere
by red supergiants.  We emphasize again that the analysis of the \mlk\
ratio of paper~I strongly supports this hypothesis for the regions where 
the CO bandheads alone do not allow us to constrain the luminosity class
of the evolved stars.

\subsection{Spatial Distribution of the Burst Ages and Strengths}
            \label{Sub-age_pixels}

Figure~\ref{fig-modpix} compares the data with single-burst models for
$m_{\rm up} = 100~{\rm M_{\odot}}$, and $t_{\rm sc} = 1$ and 5~Myr, in
the diagrams of neon, \lbol/\llyc, and \lk/\llyc\ ratios versus \ewcoa.
The size of the data points is proportional to the intrinsic \lk.  We did
not account for the contribution of the cool evolved stars to the bolometric
luminosity of individual pixels, so that the values of \lbolOB/\llyc\ are 
plotted in the \lbol/\llyc\ diagram.  

A large fraction of the data points
in all three diagrams do not fall on the model curves and exhibit the typical
behaviour encountered for the selected regions: the neon and \lbolOB/\llyc\
ratios imply significantly younger ages than \ewcoa, with \lk/\llyc\ 
corresponding to intermediate ages.  In the \lk/\llyc\ versus \ewcoa\ diagram,
the data points are distributed along a path that is remarkably parallel to
the model curve for $t_{\rm sc} = 1~{\rm Myr}$, but displaced towards lower
\lk/\llyc\ by about an order of magnitude.  
Although the properties of some of the individual pixels could be reconciled
with a single burst if the timescale is increased to values approaching
5~Myr, we have pursued with our hypothesis of sequential bursts.
The data can be reproduced with two short successive bursts, with
the young bursts producing about ten times more ionizing luminosity
than the older bursts dominating the near-IR luminosity.

For each pixel, we therefore fitted a two-burst model to the
constraints, each with $t_{\rm sc} = 1~{\rm Myr}$, in the same
manner as for the selected regions.  We assigned an age of 1~Myr for
\lbolOB/\llyc\ ratios lower than the minimum \lbol/\llyc\ predicted by 
the models (this is the case for a few pixels only).  For the old bursts,
we adopted the youngest solutions inferred from \ewcoa\ and assigned the 
age of maximum \ewcoa\ to the few pixels with EWs exceeding this limit.
We estimated the formal uncertainties on the burst ages and initial star
formation rates $R_{0}$ by varying in turn the observational constraints
in the ranges given by the measurements uncertainties.

Figure~\ref{fig-resmaps} presents the results.  The ages for the young bursts
range from 3.5 to 6.6~Myr, with an average of 4.8~Myr and dispersion of
$\sigma = 0.5~{\rm Myr}$.  The ages for the old bursts range from 7.2 to
12.6~Myr (the upper limit from the maximum \ewcoa), with an average of
8.8~Myr and dispersion of $\sigma = 0.9~{\rm Myr}$.  
Figure~\ref{fig-resprof} shows the average burst age and total burst strength
as a function of projected distance from the nucleus in a slit 3\arcsec --wide
along the galactic plane of \mqd, as indicated in Figure~\ref{fig-resmaps}.  
The typical uncertainties are $\rm \pm 0.5 - 1~Myr$ for the ages and
about 30\% for the strengths.  In view of these uncertainties,
we conclude that the maps and profiles indicate roughly constant ages
among the young bursts and among the old bursts, as well as nearly uniform
strengths for the young bursts.  Spatial variations in strength among the
old bursts are more significant with, in particular, a clear peak around
the nucleus consistent with our findings from the modeling of selected
regions (\S~\ref{Sub-resreg}).

\citet{Sat97} investigated the spatial variations in ages from a dozen compact
$K$-band emission sources, most of which are located within the regions mapped
with 3D.  They inferred burst ages in the range $\rm 4 - 10~Myr$, with a trend
of younger ages at larger projected radii.  Given that they modeled the sources
as single bursts and used the strength of the CO bandheads longwards of
2.29\,\micron\ together with the EW of the \brg\ emission line at 2.17\,\micron\
(inversely proportional to \lk/\llyc), their results are consistent with ours.

\subsection{The Global Star Formation History}  \label{Sub-global}

From our spatially detailed modeling, we reconstructed the global
star formation history within the 3D field of view by integrating
the initial star formation rate $R_{0}$ as a function of burst age
\tb\ over all individual pixels.  For sufficient sampling, we chose
age bins increasing logarithmically by 0.05~dex.
As long as the timescales assumed for all individual bursts are similar
and relatively short compared to the overall star formation history
of interest, the integrated $R_{0}(t_{\rm b})$ curve gives a measure
of the ``instantaneous'' star formation rate at different times.
Figure~\ref{fig-SFH} shows the derived integrated starburst history
together with the surface density of star formation rate, computed by
dividing the integrated $R_{0}(t_{\rm b})$ by the total area of the pixels
contributing to each age bin.  The results for the individual pixels are
reproduced as well.

The integrated star formation rate exhibits two conspicuous peaks near 5
and 9~Myr, corresponding to distinct global starburst episodes.  The first
episode was globally 2.5 times stronger than the most recent one in terms of
mass of stars formed.  The star formation rate surface density reflects the
nearly uniform strengths for the young bursts and the larger variations for
the old bursts seen in Figures~\ref{fig-resmaps} and \ref{fig-resprof}.
We note that modeling the individual pixels with two short successive bursts
introduces a bias towards separated burst events in the integrated star
formation history.  Again, such a scenario is however plausible in view
of the spatial anticorrelation between the tracers of OB stars and those
of cool evolved stars (\S~\ref{Sub-muptsc}).

The integrated star formation rate in the 3D field of view is well reproduced
by two Gaussians in $R_{0}$ versus age centered at 4.7 and 8.9~Myr, with
amplitudes of 6.3 and $18.5~{\rm M_{\odot}\,yr^{-1}}$ and full-widths
at half-maximum of 1.1 and 1.7~Myr, respectively.  These Gaussians are shown
in Figure~\ref{fig-SFH} along with the two-burst model fit to the integrated
properties of the 3D field of view from \S~\ref{Sect-models_regions}.
The exponentially decaying functions provide reasonable though crude
approximations to the detailed star formation history.
In view of our results, it may not be surprising that \citet{Rie93}
found that a double Gaussian profile optimized their model fits for the
entire starburst core (with one of their best models reaching peak star
formation activity near 5 and 13~Myr).

\section{THE NATURE OF STARBURST ACTIVITY IN M\,82}  \label{Sect-discussion}

\subsection{Star Formation Process}  \label{Sub-SFprocess}

Observations of \mqd\ reveal important small-scale structure tracing
individual burst sites on 10-pc or even parsec scales (see paper~I).
In particular, the ionized and molecular gas as well as the IR-emitting
dust exhibit clumpy morphologies on scales at least as small as 15~pc
\citep[e.g.,][]{Tel92, Lar94, Ach95, She95}.
Compact continuum sources tracing clusters of red supergiants near
their maximum luminosity are detected at near-IR wavelengths
\citep[e.g.,][]{Sat97, MCr03}.  Optical imaging with {\em HST\/} has
resolved over a hundred young super star clusters, some of which may lie in
directions of lower extinction and belong to the inner starburst regions
\citep{OCo95}.

Our models show that the properties of individual burst sites are consistent
with the formation of very massive stars and with short burst timescales of a
few million years or less \citep[see also][for selected clusters]{Sat97}.
The typical sizes and the star formation parameters inferred make these burst
sites comparable to Galactic and near-extragalactic massive star-forming
regions such as NGC\,3603 and the R\,136 cluster in 30~Doradus.

Interestingly, the smoother and low-surface brightness emission regions make
a substantial contribution to the integrated properties, suggesting that the
prominent sub-structure may in fact trace the largest and most luminous burst
sites.  The diffuse ionized gas emission (notably in H recombination lines)
and thermal dust emission represent about 50\% of the respective total emission
from the starburst core.  The smoother and fainter stellar $K$-band continuum
emission contributes at least as large a fraction to the total $K$-band
luminosity.  These components correlate roughly with the overall distribution
of the brighter and more compact sources, and partly break up into
smaller-scale structure at higher angular resolution
\citep[see, e.g., the {\em HST\/} NICMOS maps of][]{Alo03}.  Whether
the diffuse emission in \mqd, and more generally in starburst galaxies,
traces large numbers of unresolved fainter compact sources or truly extended
sources is still an open question, with implications for the importance
of cluster formation in starbursts and subsequent dissolution in the
stellar field \citep[e.g.,][]{Ho96, Meu95, Tre01}.

\subsection{Evolution of Starburst Activity}  \label{Sub-SFevol}

\subsubsection{Key Morphological Features}

For a plausible evolutionary scenario of starburst activity in \mqd,
our results must be interpreted together with the key morphological
features: \\
$\bullet$  Large-scale tails and bridges of material emanating 
from \mqd\ and connecting with its massive neighbour M\,81 located about
36~kpc in projection indicate gravitational interaction between the two
galaxies \citep*[e.g.,][]{Yun93, Yun94}. \\
$\bullet$ The morphology of the $K$-band emission, dominated 
by red supergiants, is highly suggestive of a nearly edge-on disk,
peaks strongly towards the nucleus, and provides the strongest
evidence for a $\rm \sim 1~kpc$--long stellar bar
\citep[e.g.,][]{Tel91, McL93, Lar94}.
The brightest and most compact sources are mainly located in an 
``inner disk'' within $\rm \approx 150~pc$ of the nucleus and at the
ends of the bar \citep[e.g.,][]{Sat97}.  \\
$\bullet$ The most prominent \ion{H}{2} regions are concentrated in a 
rotating ring-like structure of radius $\rm \approx 85~pc$ and, outside
this ring, along ridges presumably on the leading edge of the rotating
stellar bar \citep[e.g.,][]{Lar94, Ach95}. The ``ionized ring''
is located just inside of the main concentrations of molecular
gas in a toroid or tightly-wound spiral arms \citep[e.g.,][]{She95}.
Our \brg\ source B2 coincides in projection with the
western edge of the ionized ring while B1 is located farther out. \\
$\bullet$ The important population of young supernova remnants (SNRs)
detected at radio wavelengths is quite uniformly distributed along
the galactic plane over $\rm \approx 600~pc$
\citep*[e.g.,][]{Kro85, Mux94}.
Their spatial distribution together with the ages derived for each
starburst episode and the IMF weighting are more consistent with their
being associated primarily with the populations of red supergiants
throughout the disk. \\
$\bullet$ A bipolar outflow along the minor axis of \mqd\ is
traced out to at least 5~kpc by X-ray and optical observations 
\citep*[e.g.,][]{Bre95, Sho98}.
The [\ion{Fe}{2}] 1.644\,\micron\ emission exhibits a prominent arc-like
structure to the south and centered near the nucleus.  It may trace
shock-ionized iron-enriched gas where the outflowing wind interacts
with interstellar disk material \citep[e.g.,][]{Gre97, Alo03}. \\
$\bullet$ There is an apparent lack of gas and dust within the 
central few tens of parsecs of \mqd\ 
\citep[e.g.,][]{Tel92, Lar94, She95, Ach95, Sea96}.

Given the relative distributions of the gaseous and stellar components and
since \mqd\ is viewed at an inclination angle of $\approx 80^{\circ}$, the two
sequential bursts inferred locally and globally within the central starburst
regions could be understood in terms of projection effects.  In particular,
the inner few tens of parsecs harbored the most intense starburst activity
about 10~Myr ago and while more recent bursts appear superposed, they took
place predominantly in circumnuclear regions at larger radii.
The geometrical picture of \mqd\ further emphasizes the particular
intensity of the old nuclear burst revealed in Figures~\ref{fig-resmaps}
and \ref{fig-resprof}, with the red supergiants within the central few
tens of parsecs in projection being physically close to the nucleus
along the line of sight as well, while the OB stars --- and the red 
supergiants elsewhere --- possibly more extended along the line of sight.
While alternatives could be advanced, we chose, in line with our
two-burst hypothesis, to focus on the above interpretation as an
illustrative yet possible case.

\subsubsection{Triggering Mechanism}

The primary triggering mechanism for starburst activity in \mqd\ 
is generally attributed to the $\rm M\,82 - M\,81$ tidal interaction
$\sim 10^{8}~{\rm yr}$ ago
\citep[e.g.,][]{Got77, OCo78, Lo87, Yun93, Yun94}.
In this scenario, the ISM in \mqd\ experienced strong large-scale torques 
and loss of angular momentum as it was transported towards the dynamical
center of the galaxy, in accordance with numerical simulations 
\citep[e.g.,][]{Sun87, Nog87, Nog88, Mih96}.

The increased cloud-cloud collision rate in the disk and the large amounts of
material accumulated and compressed in the innermost regions could have led
to the first starburst episode, characterized by very intense star formation
at the nucleus and lower-level activity elsewhere in the disk.
The nuclear burst and subsequent high rate of supernova explosions consumed
rapidly the gas supply and expelled the remaining gas via the starburst wind,
thereby creating the central cavity in the ISM and preventing further star
formation.  The absence of a large concentration of radio SNRs in the nuclear
vicinity is consistent with the lack of gas and dust.

\subsubsection{Subsequent Evolution}

Numerical simulations show that bars in galactic disks can be induced
by galactic interactions and are very effective at driving material
towards the central regions of galaxies 
\citep[e.g.,][]{Com85, Nog87, Nog88, Shl89, Ath92}. 
Ring- or spiral-like dynamical resonances may develop under the action
of such non-axisymmetric perturbations.  In the presence of inner Lindblad
resonances (ILRs), the radial inflow of material is stopped before it
reaches the nucleus of the galaxy and accumulates in a circumnuclear ring.
Star formation is triggered by shocks in the ring and along ridges leading
the bar, and may be particularly enhanced at their intersections. 

The second starburst episode in our models of \mqd\ could thus be attributed
to the presence of the bar and accompanying ILRs indicated by the global gas
distributions \citep[see also][]{Lo87, Tel91, Ach95, Nei98}.  The ionized
ridges presumably trace star formation sites on the leading side of the bar.
Enhanced activity where the ISM streaming along the bar meets with the ionized
ring may be hinted at by the slight increase in strength of the young bursts
near B2 suggested by Figures~\ref{fig-resmaps} and \ref{fig-resprof}
\citep[see also][]{Ach95}.  The bar may have played a role during the
first starburst episode by channeling the nuclear inflow before the
present dynamical resonances appeared.

The apparent outward progression of
starburst activity suggested by the sequence
Nuclear supergiants $\rightarrow$ Ionized ring $\rightarrow$ Molecular ring may
reflect the temporal development of the bar and redistribution of the mass as
suggested by \citet{Tel91}.  At the radius of the ionized ring, the rotational
velocity is about $\rm 120~km\,s^{-1}$ \citep{Ach95, She95}.  This implies an
orbital period of $\rm 4 - 5~Myr$, i.e. a dynamical timescale similar to
the time elapsed between the peak star formation activity during the two
recent starburst episodes.

An alternative explanation could be self-induced propagation radially outward
in the disk, as a consequence of the powerful nuclear burst and expanding
shock wave generated by the massive stars winds and supernova explosions
\citep[e.g.,][]{Car91, Ach95}.
Without dynamical resonances, the ISM may be expected to follow a smooth, 
radially decreasing density profile.  Star formation triggered by an expanding
``superbubble'' in such a medium seems difficult to reconcile with pronounced
and narrow peaks in the global star formation rate.  The two peaks could
perhaps reflect a sudden increase of mechanical energy release by the older
nuclear burst after about 5~Myr but the contribution by massive star winds is
already very substantial at ages $\rm < 5~Myr$ before the first supernovae
explode \citep[e.g.,][]{Lei99}.  Star formation activity along the ridges
beyond the ionized ring is also difficult to explain in the self-induced
outward propagation scenario.

We favoured bar-driven evolution as dominant mechanism because the stellar
and gaseous components exhibit morphological features strongly supporting
the existence of a bar and its dynamical effects.  
Bar-driven evolution has been proposed for other starburst systems
as well as more quiescent spiral galaxies in which circumnuclear rings
of enhanced gas density and star formation activity coexist with stellar
or gaseous bars \citep*[e.g.,][]{Tel88, Tel93, Kna95, Bok97}.

\subsubsection{Global Picture}

Investigations of regions outside of the central active starburst core outline
further key elements for a global scenario.  The integrated optical spectrum
of \mqd\ exhibits strong Balmer line transitions in absorption typical of
A-type stars, indicative of an intermediate-age ($\rm 0.1 - 1~Gyr$)
``post-starburst'' population \citep{Ken92}.  \citet{Gri00, Gri01} used
{\em HST\/} optical and near-IR data to study post-starburst regions
$\rm 0.5 - 1~kpc$ northeast of the nucleus, identified as such from
strong Balmer absorption lines and discontinuity in their spectra
\citep{OCo78, Mar96}.  The results of de Grijs and coworkers based on
evolutionary synthesis modeling of the photometric properties of compact
luminous clusters and bright giant stars as well as their detection of
candidate H$\alpha$ SNRs imply that enhanced star formation occurred
$\rm \sim 600~Myr$ ago at radii of $\rm \sim 1~kpc$ and up to
$\rm \sim 30~Myr$ ago near $\rm \sim 500~pc$.  \citet{Gal99} modeled 
high-quality optical spectra of two super star clusters about 450~pc
southwest of the nucleus and derived ages of $\rm \sim 60~Myr$.

In the framework of our models and interpretation, together with the spatial
distribution of the radio SNRs and the \mlk\ ratio, the first starburst episode
$\rm \sim 10~Myr$ ago was not strictly confined to the nuclear vicinity, but
occurred throughout the central 500~pc as well.  Along with the evidence from
the post-starburst regions, this would support an ``outside-inside-out''
propagation of starburst activity in \mqd.  The scenario we propose is 
summarized here and sketched in Figure~\ref{fig-picture}.

$\bullet$ Following the gravitational interaction between \mqd\ and M\,81,
$\sim 10^{8}~{\rm yr}$ ago, the ISM in \mqd\ experienced strong large-scale
torques, loss of angular momentum, and important infall towards the nuclear 
regions, leading to enhanced star formation activity in the central 
kiloparsec.  The stellar bar induced by the interaction possibly played 
a role in channeling the inflow.  In the absence of dynamical resonances,
the infalling material can reach the nucleus.

$\bullet$ Within the central 500~pc of \mqd, a first starburst episode took
place $8 - 15~{\rm Myr}$ ago and was rapidly exhausted.  The inner few tens
of parsecs at the nucleus hosted the most intense star formation activity.

$\bullet$ A subsequent starburst episode was triggered predominantly
by bar-induced dynamical resonances.  It occurred $\rm 4 - 6~Myr$
ago, mainly in a circumnuclear ring and along the stellar bar, and 
also decayed rapidly.

$\bullet$ A supernova-driven starburst wind originating in the center
of \mqd\ has broken out of the galactic plane, the dramatic aftermath
of the powerful nuclear burst.  The outflow component in the disk may
have played some role in triggering the second starburst episode.

\subsection{Timescales and Feedback Effects}

The global durations of $\rm \sim 10^{6}~yr$ derived for each of the recent
starburst episodes in \mqd\ seem surprising for spatial scales up to at least
500~pc and in view of the longer estimates $\rm \sim 10^{7} - 10^{8}~yr$
often quoted for starburst systems \citep[e.g.,][]{Thr86, Hec98}.
We consider below three timescale arguments in trying to gain more insight
into the quenching mechanisms of the recent starbursts in \mqd.

One argument relies on the comparison between the present star
formation rate and the mass of the current gas reservoir, giving a
``gas consumption timescale'' $\tau_{\rm gas}$.
In \mqd, the gas is mostly in molecular form, with a total
$M_{\rm H_{2}} = 1.8 \times 10^{8}~{\rm M_{\odot}}$ 
\citep[e.g.,][]{Wil92}.  From our two-burst model of the starburst
core (Table~\ref{tab-modglob}) and with a Salpeter IMF, 
$M^{\star}_{\rm stb} \approx 2 \times 10^{8}~{\rm M_{\odot}}$ of stars were
formed in the past $\rm \approx 15~Myr$ for $m_{\rm low} = 1~{\rm M_{\odot}}$
and $\approx 5 \times 10^{8}~{\rm M_{\odot}}$ for
$m_{\rm low} = 0.1~{\rm M_{\odot}}$.
The latter value represents 80\% of the measured stellar mass and, in view of
the discussion of \S~\ref{Sub-mlow}, is here considered as an upper limit.
The average star formation rate in the recent history of \mqd\ is then
$R = 13 - 33~{\rm M_{\odot}\,yr^{-1}}$, depending on \mlow.  Assuming 100\%
efficiency, starburst activity in \mqd\ at the level observed in the past
15~Myr could be sustained for another $\rm \tau_{\rm gas} \approx 5 - 15~Myr$. 
This provides a fair estimate for the minimum gas depletion timescale but has
no connection with the physical processes responsible for the evolution of
the starburst.  

Considerations based on dynamical timescales are more closely linked
to the large-scale mechanisms that may drive starburst activity.
For a sample of 36 IR-selected starburst galaxies, \citet{Ken98}
inferred a median gas conversion efficiency of 30\% per dynamical
timescale, taken as an average orbital period for all objects of
$\rm 10^{8}~yr$; $\epsilon_{\rm gas} = 100\%/10^{8}~{\rm yr}$ occurs
but in the most extreme objects.  From the rotation curve of \mqd\
\citep[e.g.,][]{Got90, McK93, Ach95, She95, Nei98}, we compute shorter
dynamical timescales for regions inside the starburst core.  For example, the
orbital periods are about 5, 10, and 20~Myr at radii of 90, 225, and 500~pc
corresponding to the locations of the ionized ring, of the main concentrations 
of molecular gas, and of the outer limit enclosing the starburst core and most 
of the gas.  The dynamics allow then for burst timescales of approximately
$\tau_{\rm dyn} = 15~{\rm Myr}$, $\rm 30~Myr$, and $\rm 65~Myr$
for 30\% efficiency per orbital period.  Even with 100\% efficiency, 
the timescales are barely short enough to explain the durations of
each starburst episode in \mqd.

Such dynamical considerations may provide more insight concerning the
settling time of the gas and the development of dynamical resonances.
The orbital time at the radius of the ionized ring is
similar to the time separation between the two most recent bursts.
At radii $\rm 1 - 3~kpc$, the orbital period is $\rm \approx 100 - 200~Myr$,
which is of the order of the age of the super star clusters in the
post-starburst regions $\rm \sim 1~kpc$ northeast of the nucleus and
to the epoch of last peri-passage between \mqd\ and M\,81.  The scenario
proposed in \S~\ref{Sub-SFevol} for the triggering and propagation of
the starburst episodes in \mqd\ is consistent with these timescales.

None of the above arguments is however completely satisfactory for
explaining the very short global burst durations suggested by our models.
Most importantly, any feedback effect of massive stars via strong stellar
winds and subsequent supernova explosions is neglected.
Qualitatively, timescales of a few million years are similar to the 
lifetimes of stars with initial masses $\rm \gtrsim 50~M_{\odot}$.
Although both episodes took place over $\rm \sim 100 - 500~pc$ scales,
each was triggered nearly simultaneously everywhere, presumably as a result
of brief gas compression events.  Massive star winds and supernova explosions
rapidly disrupting the remaining ISM locally, preventing further star formation
after a few million years, may have ``conspired'' to produce a rapid decay
of the global starburst activity as well.

We can make simple estimates for the ``feedback timescale'' $\tau_{\rm feed}$
by comparing the cumulative mechanical energy $E_{\rm mech}$ injected by
supernovae (and neglecting the contribution by stellar winds) into the ISM
with its gravitational binding energy $E_{\rm bind}$, and asssuming that
starburst activity stops when they exactly balance each other:
\begin{equation}
\dot{E}_{\rm mech} \tau_{\rm feed} = E_{\rm bind},
\label{Eq-cond}
\end{equation}
where $\dot{E}_{\rm mech}$ is the rate of mechanical energy deposition.
We assumed that all stars with initial masses $\rm \geq 8~M_{\odot}$ explode
as supernovae with a typical energy of $\rm 10^{51}~erg$, of which
a fraction $\eta$ is transferred as kinetic energy to the ISM.
For simplicity, we considered a spherically symmetric distribution of 
gas and stars within the starburst core, so that 
$E_{\rm bind} = \left(G M_{\rm dyn} M_{\rm gas}\right)/{\mathcal{R}}$
where $G$ is the gravitational constant, $M_{\rm dyn}$ and $M_{\rm gas}$ are
the dynamical and gas masses, and $\mathcal{R}$ is the radius of the region.

The rate of supernova explosions \snrate\ is the most direct tracer of 
$\dot{E}_{\rm mech}$ but
the observed rates suffer from rather large uncertainties.  We computed
$\dot{E}_{\rm mech}$ using the inferred star formation rate $R$ instead.
The implicit assumption is that the burst is sufficiently evolved to have
non-negligible \snrate\ and that the star formation rate is still substantial.
For a Salpeter IMF between 1\masssol\ and 100\masssol,
$\nu_{\rm SN}~[{\rm yr^{-1}}] \approx 0.02\,R~[{\rm M_{\odot}\,yr^{-1}}]$.
With these assumptions, Eq.\,\ref{Eq-cond} can be written: 
\begin{equation}
\frac{\tau_{\rm feed}}{\rm Myr} \simeq 500\left(\frac{f}{\eta}\right)
\left(\frac{M_{\rm dyn}}{\rm 10^{10}~M_{\odot}}\right)^{2}
\left(\frac{\mathcal{R}}{\rm kpc}\right)^{-1}
\left(\frac{R}{\rm M_{\odot}\,yr^{-1}}\right)^{-1},
\label{Eq-SFR}
\end{equation}
where $f$ is the gas mass fraction.  For the starburst core of \mqd,
$\mathcal{R} = 225~{\rm pc}$,
$M_{\rm dyn} \approx 8 \times 10^{8}~{\rm M_{\odot}}$, and
$M_{\rm gas} \approx 1.8 \times 10^{8}~{\rm M_{\odot}}$ (see paper~I).
With the average $R \approx 13~{\rm M_{\odot}\,yr^{-1}}$ derived for the
appropriate IMF parameters, Eq.\,\ref{Eq-SFR} yields
$\tau_{\rm feed} \simeq 0.2 \eta^{-1}~{\rm Myr}$.  The supernova
efficiency is poorly constrained, but values in the range $\eta = 0.1 - 1$
seem appropriate for \mqd\ \citep{Che85}.  Consequently,
$\tau_{\rm feed} \simeq 0.2 - 2~{\rm Myr}$. 
For comparison, with $\nu_{\rm SN} \approx 0.06~{\rm yr^{-1}}$ for \mqd, 
we obtain $\tau_{\rm feed} \simeq 1 - 10~{\rm Myr}$ depending on $\eta$.
As alternative tracer, we can use \lbol.  In a steady-state, our starburst
models imply 
$L_{\rm bol}~[{\rm L_{\odot}}] \sim 10^{12}\nu_{\rm SN}~[{\rm yr^{-1}}]$, 
and thus $R~[{\rm M_{\odot}\,yr^{-1}}] \sim 
0.5\left(L_{\rm bol}/10^{10}~{\rm L_{\odot}}\right)$.
With $L_{\rm bol} = 6.6 \times 10^{10}~{\rm L_{\odot}}$ for the starburst core,
Eq.\,\ref{Eq-SFR} implies $\tau_{\rm feed} \sim 1 - 10~{\rm Myr}$ as well.
These estimates represent upper limits because we neglected
massive stars winds.

The $\tau_{\rm feed}$ is more consistent than $\tau_{\rm gas}$ or 
$\tau_{\rm dyn}$ with the short durations in our models for the recent
starburst episodes in \mqd.  Admittedly, our $\tau_{\rm feed}$ estimates
are also simplistic but they provide an additional perspective for the
evolution of starburst activity and a possible explanation for short global
burst durations.  Along with our spatially detailed models, the above
timescales are consistent with the overall progression of starburst activity
in \mqd\ being driven by large-scale dynamical processes related to the tidal
interaction with M\,81 and the induced stellar bar, but suggests that once
starburst activity was triggered, strong negative feedback effects acted
very rapidly to inhibit further star formation on all spatial scales.

Recent studies of other starburst systems also provide evidence for
short burst durations and for the episodic nature of starburst activity
\citep[e.g.,][]{Tho00, Bok97, Van98, Alo01}.
Some numerical simulations of the dynamical evolution of barred galaxies 
which account for the effects of star formation indicate a recurrent burst
behaviour with typical timescale $\rm \sim 10^{7}~yr$, owing to the ability
of massive stars to destroy the flow pattern when a sufficient number
coexist in a given place \citep{Hel94, Kna95}.

\section{SUMMARY}  \label{Sect-conclu}

We have presented the results of detailed modeling of the central
starburst regions of \mqd, based on new near-IR integral field
spectroscopy and mid-IR spectroscopy complemented with additional
results taken from the literature. 
We applied starburst models optimized for observations in the
range $\rm \lambda = 1 - 45~\mu m$ to these data and constrained the star
formation parameters (cutoffs of the IMF, burst timescale) as well as the
spatial and temporal evolution of starburst activity on scales as
small as 25~pc.

Our data and models are consistent with the formation of very massive stars
($\rm \gtrsim 50~M_{\odot}$) and imply burst durations of at most a few
million years for individual burst sites.  With typical sizes on 10--pc
scales or less, individual burst sites in \mqd\ are comparable to Galactic
and near-extragalactic massive star-forming regions such as NGC\,3603 and
the R\,136 cluster in 30~Doradus.
For plausible limits on the mass involved in the starburst, we find that
the IMF must flatten at low masses assuming a Salpeter slope at high masses.
Dynamical information for three individual young stellar clusters in \mqd\
\citep{Smi01, MCr03} provide similar evidence, but clearly more clusters need
to be investigated.  In view of the large uncertainties involved and the
scarcity of low-mass IMF determinations in local high-mass star-forming
regions, this result is difficult to interpret.

Among possible alternatives, our model results and interpretation
together with the spatial distribution of the stellar and gaseous
components in \mqd\ leads us to propose the following scenario.
Globally, starburst activity in the central 500~pc of \mqd\ occurred
during two successive episodes each lasting a few million years, about
10 and 5~Myr ago.  The first episode took place throughout the disk
and was particularly intense in the central few tens of parsecs.  The
second episode occurred predominantly in circumnuclear regions and
along the stellar bar.  We interpret this sequence in a tidally-triggered,
bar-driven evolution scenario, consistent with the evidence for
gravitational interaction between \mqd\ and its neighbour M\,81, 
with the observed morphological signatures of the bar and associated
dynamical resonances, and with the corresponding dynamical timescales.

The episodic nature of starburst activity and short global burst decay
timescales in \mqd\ are particularly interesting results.
Invoking a simple argument based on the comparison of the cumulative
mechanical energy injected into the ISM and its gravitational binding energy,
we find that the collective effect of massive stars winds and supernova
explosions can disrupt the ISM in \mqd\ on timescales of
$\rm \sim 10^{6}~yr$, providing a very efficient quenching mechanism
for each starburst episode.
Following brief gas compression events on large-scales, starburst activity
decayed very rapidly due to its own strong negative feedback effects.
Our scenario for \mqd\ outlines the interplay between the large-scale
triggering mechanisms and the more local but important feedback effects 
in determining the evolution of starburst activity in this galaxy.

\acknowledgments
We would like to thank A. Pauldrach and R.-P. Kudritzki for providing us
with their model atmospheres, G. Ferland for making CLOUDY version C90.05
available in advance of publication, and C. Telesco and D. Gezari for their
12.4\,\micron\ data.
It is a pleasure to thank M. Thornley, M. Lehnert, J. Gallimore, L. Tacconi, 
and L. Tacconi-Garman for stimulating discussions and useful comments, 
and to H. Spoon for help with using CLOUDY.  
We also wish to thank the referee, C. Leitherer, for a very considered and
helpful report.
This work was part of the Ph.D. thesis of NMFS, who
acknowledges the Fonds pour les Chercheurs et l'Aide \`a la Recherche
(Gouvernement du Qu\'ebec, Canada) for a Graduate Scholarship,
and the Max-Planck-Institut f\"ur extraterrestrische Physik and
Service d'Astrophysique of the CEA Saclay for additional financial support.
SWS and the {\em ISO\/} Spectrometer Data Center at MPE are supported
by DLR under grants 50 QI 8610 8 and 50 QI 9402 3.
We also thank the German-Israeli Foundation for support of this work
(grant I-551-186.07/97).



\appendix

\section{FURTHER DETAILS ON STARS}   \label{App-STARS}

\subsection{Spectral Energy Distributions}

STARS computes the detailed SED intended for photoionization modeling of the
nebular emission excited by the stars.  The SEDs are taken from a hybrid grid
generated as described by \citet{Tho00} from the LTE models of \citet{Kur92}
for effective temperatures $T_{\rm eff} \leq 19,000~{\rm K}$ and the
non-LTE models of \citet{Pau98} for $T_{\rm eff} \geq 25,000~{\rm K}$,
with SEDs for intermediate temperatures obtained by interpolation.
The \citeauthor{Pau98} models represent much better the effects of line
blocking and blanketing in the rapidly expanding atmospheres of hot stars
and are generally harder than the \citeauthor{Kur92} models.  A detailed
description of the models and implications on the computed SEDs is given
by \citet{Pau01} and references therein.  The basic \citeauthor{Pau98}
model SEDs we used in STARS include dwarfs and supergiants.
For dwarfs, they cover $\rm 25000 - 60000~K$ in $T_{\rm eff}$ at fixed
surface gravity $\log(g) = 4$ and mass-loss rates $\dot{M}$ of
$10^{-8} - 5 \times 10^{-6}~{\rm M_{\odot}\,yr^{-1}}$.
The ranges for supergiants are $T_{\rm eff} = 25000 - 50000~{\rm K}$,
$\log(g) = 2.75 - 3.8$, and $\dot{M}$ of
$2 \times 10^{-6} - 1.5 \times 10^{-5}~{\rm M_{\odot}\,yr^{-1}}$.
For a given $T_{\rm eff}$ and $L_{\rm bol}$ along a stellar track, STARS
uses the corresponding evolutionary $\log(g)$ and assigns the SED model with
closest $T_{\rm eff}$ and $\log(g)$ from the interpolated hybrid grid.
The SED model grid covers well the range in parameters of the Geneva
stellar tracks during most of the hot stars lifetimes (for both normal 
and enhanced mass-loss rates sets).  

Various quantities predicted by STARS are derived using the empirical
bolometric corrections of \citet{SK82} and the broad-band colours of
\citet{Koo83}.  This is especially important for cool stars which have SEDs
severely distorted by extensive molecular absorption and contaminated by
thermal emission from a circumstellar dust shell that are not reproduced
satisfactorily by black-body approximations or currently available model
atmospheres.  The impact of using empirical data is largest at near-IR
wavelengths.  For instance, the predicted $K$-band luminosities are up to
30\% larger for the age range $\rm \sim 10 - 100~Myr$
(assuming $t_{\rm sc} = 1~{\rm Myr}$).

\subsection{Near-IR Stellar Absorption Features}

STARS predicts the EWs of near-IR stellar absorption features
characteristic of late-type stars: the $\rm ^{12}CO$\,(6,3) and
$\rm ^{12}CO$\,(2,0) bandheads at 1.62 and 2.29\,\micron, and the \ion{Si}{1},
\ion{Na}{1}, and \ion{Ca}{1} features at 1.59, 2.21, and 2.26\,\micron. We
implemented in STARS the EWs versus \teff\ relationships for luminosity
classes V, III, and I below 5700~K obtained by polynomial fitting to the
homogeneized stellar data compiled by \citet{FS00}.  The EWs assigned to
hotter stars were identically zero since the corresponding features vanish
at higher \teff.  The exception is the 1.59\,\micron\ feature but above
$\rm \approx 6000~K$, the hydrogen Br14 transition is responsible for
the absorption feature which has then a different physical meaning.
The predicted EWs include ``dilution'' effects due to featureless continuum
emission from OB stars and from associated free-free and free-bound processes.
The stellar data are given for spectral resolutions of $R \sim 1600$ and 2500
in the $H$ and $K$ band, respectively.  For comparison with the models, EWs
measured from lower resolution data should be corrected according to the
relationships given by \citet{FS00}.

Recently, \citet{Ori99} and \citet{Ori00} questioned the reliability of
synthesis modeling of the CO bandheads from a critical analysis of theoretical
modeling and observational data.  They emphasized the importance of accounting
for AGB evolution up to the end of the thermally-pulsing phase.  They also
demonstrated that uncertainties in stellar evolutionary tracks for red
supergiant and AGB phases remain large especially at sub-solar metallicities
but affect much less the results for ages up to $\rm \sim 10^{7}~yr$ at
near-solar metallicities.  Since STARS accounts for the thermally-pulsing AGB
(see below) and with the solar-metallicity tracks adopted in this work, these
uncertainties are minimized in our predictions of the CO EWs.

\subsection{Implementation of the TP-AGB Phase}

The Geneva tracks \citep{Sch92} employed in STARS follow the evolution of
stars up to well defined evolutionary phases.  Tracks for massive stars
($M > 7~{\rm M_{\odot}}$) extend up to the end of the core carbon-burning
phase, just prior to their explosion as core-collapse supernovae.  Tracks for
low-mass stars ($M < 2~{\rm M_{\odot}}$) end at the helium flash which occurs
at ages $\rm \gtrsim 10^{9}~yr$.  The subsequent evolution of low-mass stars
is not relevant for the starburst ages ($\rm < 10^{9}~yr$) we consider.
For intermediate-mass stars ($2~{\rm M_{\odot}} < M < 7~{\rm M_{\odot}}$),
the Geneva tracks are terminated at the end of the early asymptotic giant
branch (E-AGB), at ages in the range $\rm 5 \times 10^{7}$ to $\rm 10^{9}~yr$,
and thus exclude the TP-AGB phase.
TP-AGB stars can contribute significantly to the near-IR and total
luminosities in starbursts, and so should be included in synthesis models
\citep[e.g.,][]{CB91, Maras98, Ori00, Mou02}.
Observationally, there is growing evidence that such intermediate-age
populations account for a non-negligible fraction of the integrated light in
the central regions or bulges of several nearby galaxies, including our own
\citep*[e.g.,][]{Fre92, Blu96, McL96, Lee96, Dav97, Mar98}.

We extended the Geneva tracks for intermediate-mass stars using an
approach suggested by \citet{CB91} which employs the analysis of the TP-AGB
presented by \citet{Bed88}.  In this procedure, the TP-AGB is divided into a
beginning ``Mira phase'' where the stars pulsate in the first overtone
(or higher pulsation modes) while losing mass slowly, and a late ``OH/IR''
phase where the stars pulsate in the fundamental mode and lose mass rapidly.
In the Mira phase, the stars increase in luminosity (and cool) and move
up the AGB track appropriate for their mass.  In the OH/IR phase, the
luminosities do not vary much because of the rapid mass loss.  \citet{Bed88}
provides analysis and computations of the durations of each phase, including
the maximum luminosities $L_{\rm max}$ at the end of the OH/IR phase, as
functions of the main-sequence masses between 2 and 7\masssol.
Using this information, we added two evolutionary points in the \lbol\
versus \teff\ Hertzsprung-Russell diagram.
The first point corresponds to the end of the Mira phase where (following
\citeauthor{Bed88}) we assume that $L_{\rm bol} = 0.8\,L_{\rm max}$.
The second point corresponds to the end of the OH/IR phase where
$L_{\rm bol} = L_{\rm max}$.  We obtained the effective temperature at each
point from the $L_{\rm bol} - T_{\rm eff}$ relations for stars on the AGB.
Our extensions are illustrated in Figure~\ref{fig-TPAGB}.
As sanity check, we compared the \lbol\ predicted by our shortest burst models
based on conventional synthesis and by applying the fuel consumption theorem
\citep{Ren86} to our library of tracks.  We obtain similar values within
10\% on average up to $\rm 10^{9}~yr$, with maximum differences of $\pm 25\%$
at $\rm \leq 3 \times 10^{8}~yr$ increasing to $\pm 40\%$ at older ages where
the numerical fluctuations due to the finite number of tracks become
more important.

In attributing near-IR properties to TP-AGB stars, we followed a simple
approach which can be justified by the small dispersion in \teff\ of TP-AGB
stars of various masses and the fact that the time spent in the OH/IR phase
is at most 30\% of the total TP-AGB lifetime according to \citeauthor{Bed88}'s
models \citep[see also][]{CB91}.
We assigned to all TP-AGB stars the average bolometric correction and
broad-band colours of the prototypical Mira variable $\omega$ Cet, derived
from the data of \citet{Men65} and \citet{Zho84}.  For the EWs of near-IR
absorption features, we adopted preliminary data obtained with the 3D
instrument for a sample of Miras and other TP-AGB stars (S- and N-type stars).
The observed CO bandhead strengths are comparable to those of the latest
normal M-type giants.  Similar conclusions are reached by inspection of the
spectra published, e.g., by \citet{Lan99} and \citet{Lan00}.  We therefore
attributed the EWs of normal giants with $T_{\rm eff} = 3000~{\rm K}$
to all TP-AGB stars.

As expected, the effects of including the TP-AGB phase are most important
for \lbol\ and for the near-IR properties while they are negligible
for optical and UV properties.  Figure~\ref{fig-TPAGBmod} shows the TP-AGB
contribution in our synthetized \lbol, \lk, and \ewcoa.  Detailed comparison
with other models including the TP-AGB, notably by \citet{CB91}, 
\citet{Maras98}, and \citet{Mou02}, is difficult because of the differences in
synthesis techniques (conventional, isochrone, fuel consumption theorem) and
in assumptions for TP-AGB stars (e.g., \teff\ scale, bolometric corrections,
detailed evolution, lifetimes) among the various works.  The time evolution
and difference between \lbol\ and \lk\ of the TP-AGB contribution in our models
agree qualitatively with these other models but the comparison suggests that
the peak at $\rm \approx 5 \times 10^{8}~yr$ for $t_{\rm sc} = 1~{\rm Myr}$
may be too narrow and with too large an amplitude.  Around $\rm 10^{8}~yr$,
our short burst predictions lie in the ranges computed for coeval populations
by the above authors, from a few to $\sim 15\%$ for \lbol\ and $\lesssim 10\%$
to $\sim 40\%$ for \lk\ (with fairly large differences between different work).
We point out that we implemented the TP-AGB to assess to first order the
effects on the properties synthetized by STARS, and that our results of
\mqd\ would be essentially unchanged without it.  Accurate modeling
would require a more careful and detailed treatment as presented by, e.g.,
\citet{Maras98} and \citet{Mou02}, as well as a better understanding of the
properties and evolution of TP-AGB stars.

\clearpage


\setcounter{figure}{0}

\begin{figure}[p]
\epsscale{1.5}
\plotone{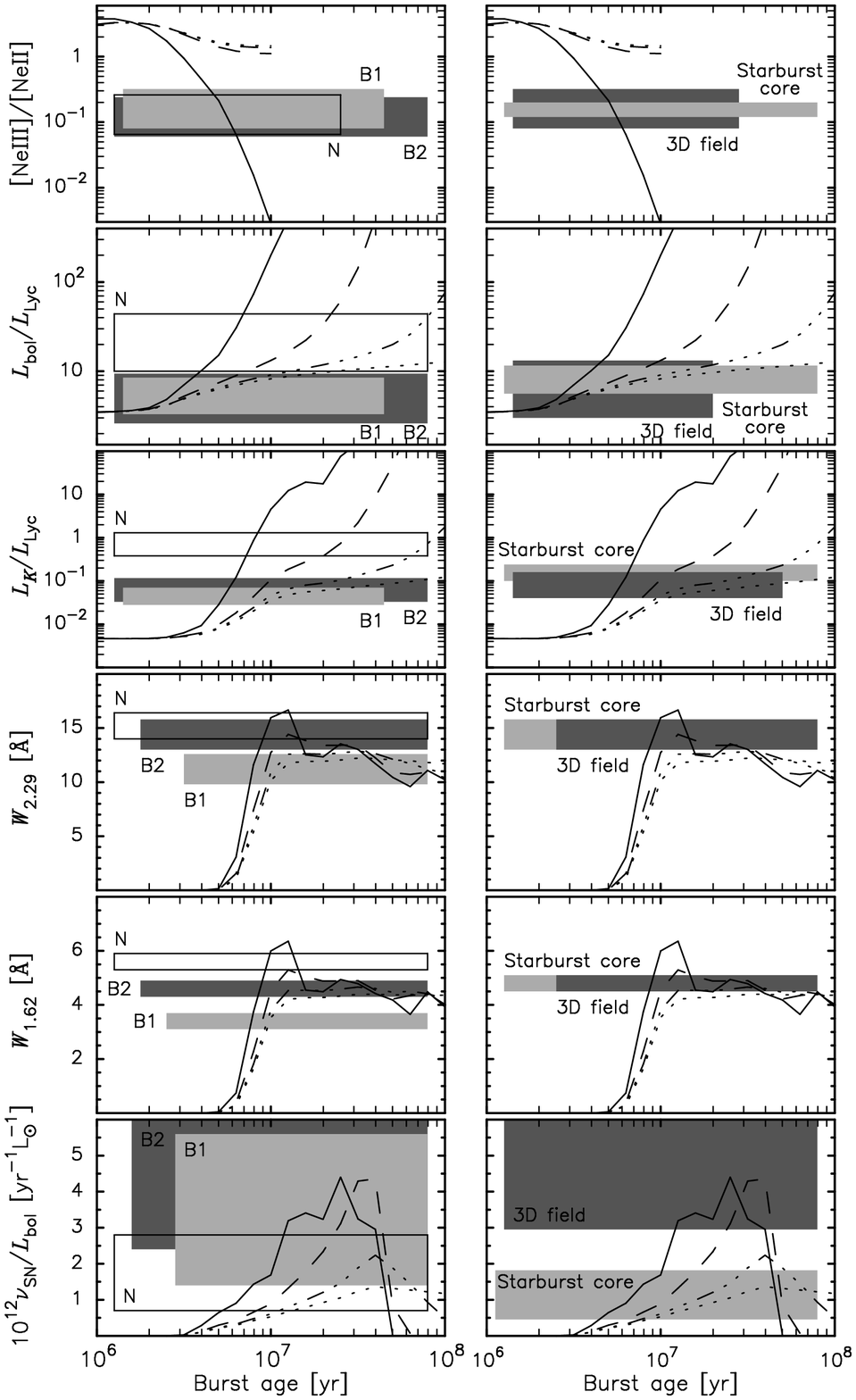}
\vspace{-1.0cm}
\caption{
Comparison of the observed properties of selected regions in
M\,82 with model predictions.  The curves are computed for
$m_{\rm up} = 100~{\rm M_{\odot}}$ and four different burst
decay timescales: 1~Myr (solid lines), 5~Myr (dashed lines),
20~Myr (dash-triple dot lines), and 1~Gyr (dotted lines).
The other model parameters are given in Table~\ref{tab-param}.
The horizontal bars indicate the measurements for the different
regions as follows.  In the left-hand side panels: central 35~pc
at the nucleus (empty bars labeled ``N''), B1 (light-shaded bars),
and B2 (dark-shaded bars).  In the right-hand side panels:
the 3D field of view and the entire starburst core
(dark- and light-shaded bars, respectively).
The width of each bar corresponds to the formal uncertainties.
\label{fig-mod_tsctb}
}
\end{figure}

\clearpage

\begin{figure}[p]
\epsscale{1.5}
\plotone{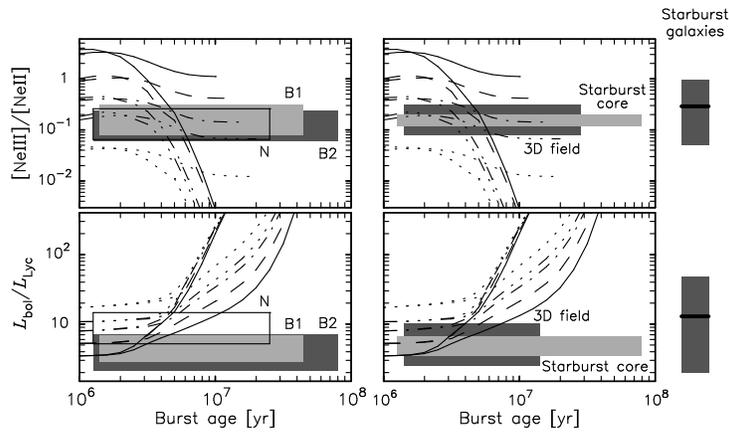}
\vspace{-2.0cm}
\caption{
Comparison of the neon and $L_{\rm bol}^{\rm OB}/L_{\rm Lyc}$ ratios
for selected regions in M\,82 with model predictions.  The curves
correspond to $m_{\rm up} = 100~{\rm M_{\odot}}$ (solid lines),
$\rm 50~M_{\odot}$ (dashed lines), $\rm 35~M_{\odot}$ (dash-dotted lines),
$\rm 30~M_{\odot}$ (dash-triple dot lines), and $\rm 25~M_{\odot}$
(dotted lines); two burst timescales are illustrated:
1 and 5~Myr (steepest and shallowest curves, respectively).
The horizontal bars within each plot indicate measurements for
different regions as in Figure~\ref{fig-mod_tsctb}.
The vertical bars and thick lines on the right-hand side of the figure
show the ranges of and average ratios for the sample of solar-metallicity
starburst galaxies observed with {\em ISO\/}-SWS by \citet{Tho00}.
\label{fig-mod_mup}
}
\end{figure}

\clearpage

\begin{figure}[p]
\epsscale{2.0}
\plotone{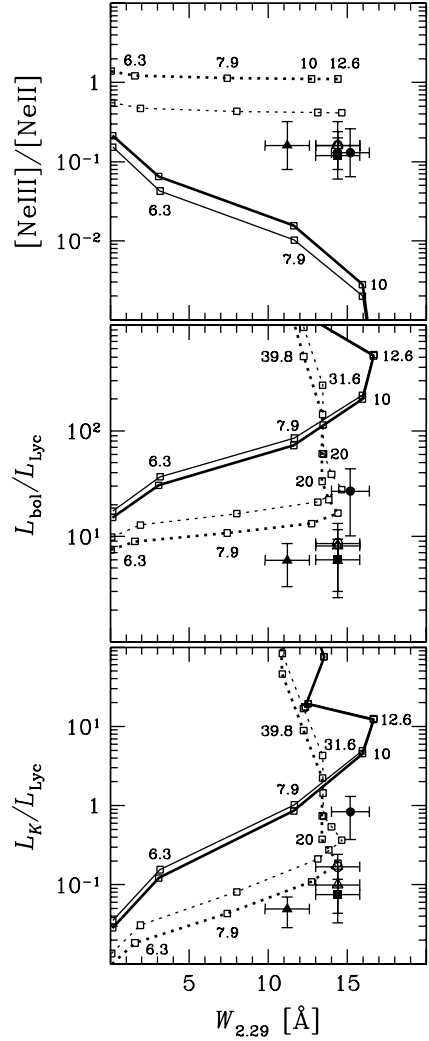}
\vspace{-1.0cm}
\caption{
Single burst evolutionary tracks in diagrams of the neon, 
$L_{\rm bol}/L_{\rm Lyc}$, and $L_{K}/L_{\rm Lyc}$ ratios versus $W_{2.29}$.
The data points with error bars indicate the measurements for the nucleus
(filled circles), B1 (filled triangles), B2 (filled squares), the 3D
field of view (open triangles), and the starburst core (open circles).
The thick and thin curves correspond to models with $m_{\rm up} = 100$
and $50~{\rm M_{\odot}}$, respectively.  Two burst decay timescales are
plotted for each $m_{\rm up}$: 1 and 5~Myr (solid and dotted line pairs,
respectively).  The open squares along each curve indicate burst ages
separated by logarithmic intervals of
$\Delta \log (t_{\rm b}\,[{\rm yr}]) = 0.1~{\rm dex}$,
with selected ages (in Myr) labeled for reference.
\label{fig-modreg}
}
\end{figure}

\clearpage

\begin{figure}[p]
\epsscale{1.9}
\plotone{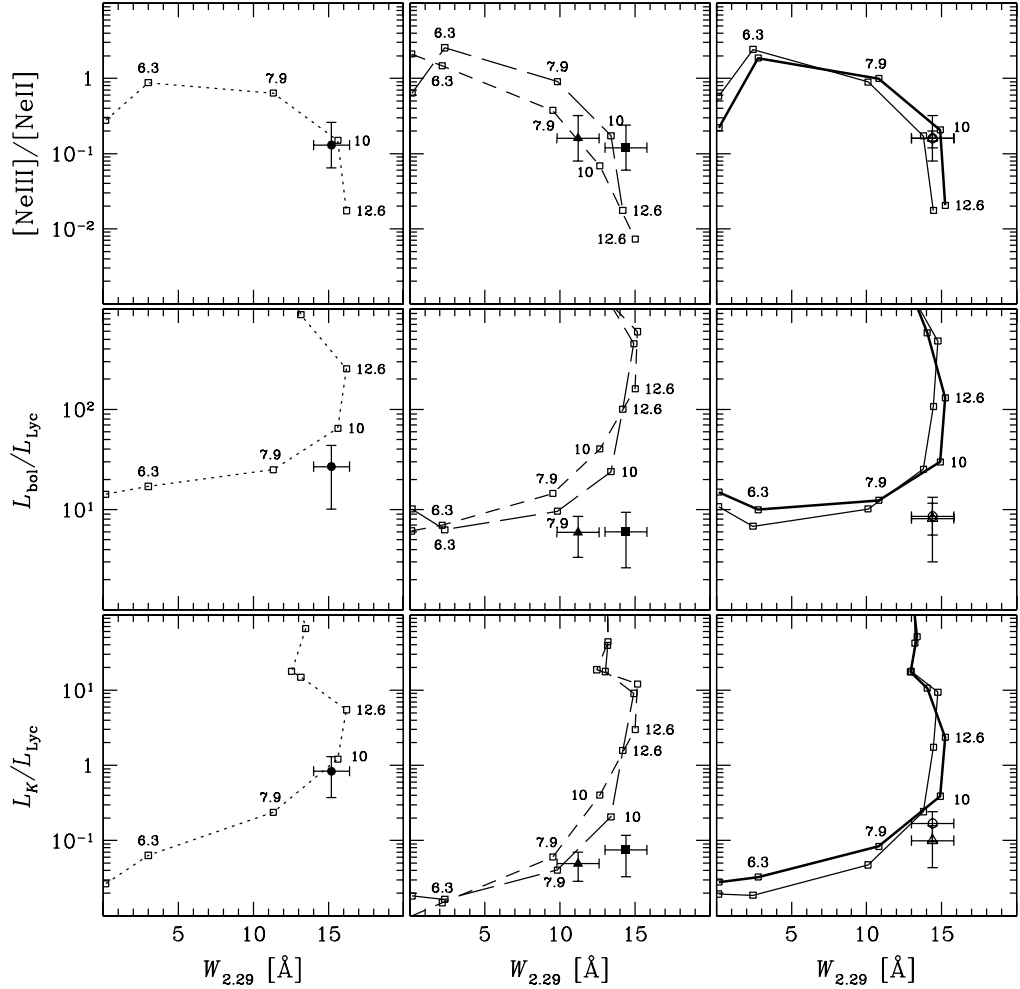}
\vspace{-1.0cm}
\caption{
Two-burst model evolutionary tracks in diagrams of the neon, 
$L_{\rm bol}/L_{\rm Lyc}$,
and $L_{K}/L_{\rm Lyc}$ ratios versus $W_{2.29}$.
The data points with error bars indicate the measurements for the nucleus
(left panels), B1 and B2 (center panels), and the 3D field of view and
starburst core (right panels).
The curves show the evolution of the properties for the
combination of two bursts with best-fit time separation and relative
intensities for each region (\S~\ref{Sub-resreg}), each burst with
$t_{\rm sc} = 1~{\rm Myr}$ and $m_{\rm up} = 100~{\rm M_{\odot}}$.
Symbols and lines are as follows:
filled circle and dotted line for the nucleus, filled triangle
and short-dashed line for B1, filled square and long-dashed line for B2,
open triangle and thin solid line for the 3D field of view, and
open circle and thick solid line for the starburst core.
The open squares and labels along each curve indicate the time in Myr
since the onset of the older burst, in logarithmic intervals of
$\Delta \log (t_{\rm b}\,[{\rm yr}]) = 0.1~{\rm dex}$.
\label{fig-modregtwo}
}
\end{figure}

\clearpage

\begin{figure}[p]
\epsscale{1.5}
\plotone{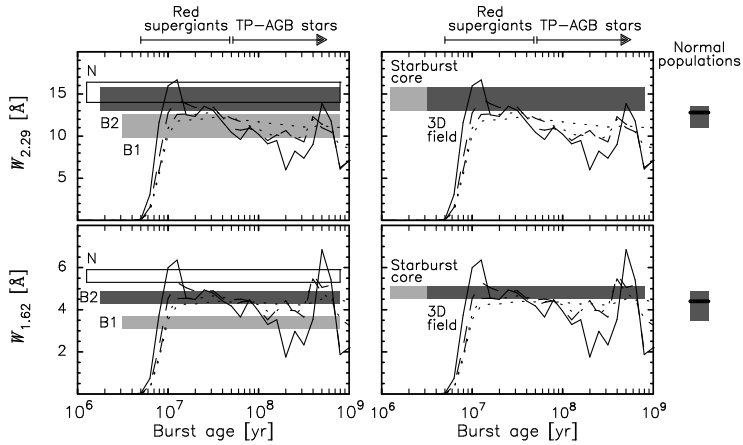}
\vspace{-3.5cm}
\caption{
Comparison of the CO bandheads EWs of selected regions in M\,82 with
model predictions.  The data and model curves are the same as in
Figure~\ref{fig-mod_tsctb}.
A larger range of burst ages is shown to compare the EWs during the
phases when red supergiants dominate the near-IR continuum and when
intermediate-mass stars become important contributors, in particular
TP-AGB stars.
The vertical bars and thick lines on the right-hand side of the figure
indicate the ranges of and average EWs observed for normal stellar
populations characteristic of elliptical galaxies and bulges of
spiral galaxies \citep{OOKM95}.
The sharp increase in synthetized EWs for $t_{\rm sc} \leq 20~{\rm Myr}$
around 500~Myr is most likely exaggerated due to the synthesis technique
employed (see appendix~\ref{App-STARS}).  Corresponding solutions are
thus not considered valid and are also not supported by the
mass-to-$K$-band light ratio.
\label{fig-EWs}
}
\end{figure}

\clearpage

\begin{figure}[p]
\epsscale{1.5}
\plotone{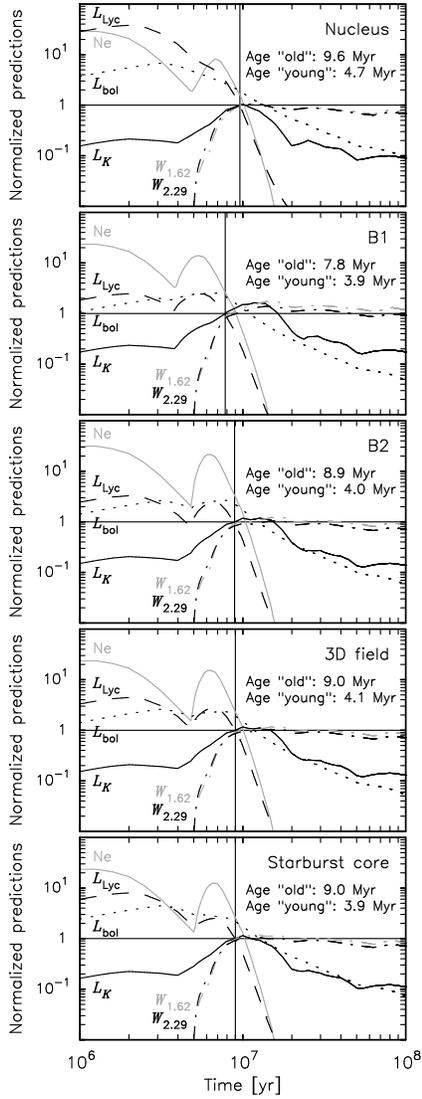}
\vspace{-0.5cm}
\caption{
Starburst models for selected regions in M\,82.  
The curves in each plot represent the evolution of the properties
for the two-burst model for each region (\S~\ref{Sub-resreg})
as a function of the time elapsed since the onset of the oldest burst:
neon ratio (grey solid lines labeled ``Ne''),
$L_{\rm bol}$ (black dotted lines), $L_{\rm Lyc}$ (black dashed lines),
$L_{K}$ (black solid lines), and $W_{1.62}$ and $W_{2.29}$
(grey and black dash-dotted lines, respectively).
The ages for the bursts (``old'' and ``young'') are given in each diagram.
The curves are normalized to the observed values; ideally, they should
all meet at unity (horizontal line) at the appropriate age for the old
burst (vertical line).
An $m_{\rm up} = 100~{\rm M_{\odot}}$ and $t_{\rm sc} = 1~{\rm Myr}$
were adopted for each burst.
\label{fig-resreg}
}
\end{figure}

\clearpage

\begin{figure}[p]
\epsscale{1.7}
\plotone{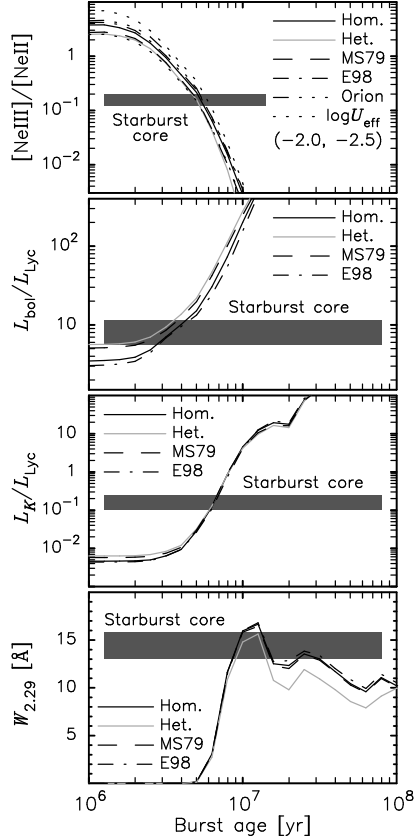}
\vspace{-3.0cm}
\caption{
Effects of different model assumptions and parameters on the
predictions of selected properties.
For all models shown, $m_{\rm up} = 100~{\rm M_{\odot}}$ and
$t_{\rm sc} = 1~{\rm Myr}$.
The black solid curves represent the models applied to M\,82, computed
assuming a homogeneous cluster population.  The other curves show the
effects of accounting for a heterogeneous cluster population following
a plausible luminosity function (grey solid lines), and of adopting the
IMF of \citet[][dashed lines]{MS79} or \citet[][dash-dotted lines]{Eis98}.
Additional curves for the neon ratio were computed for a gas and dust
composition typical of the Orion nebula (dashed-triple dot line) and for
$\log U_{\rm eff} = -2~{\rm dex}$ and $\rm -2.5~dex$ (upper and lower
dotted lines, respectively).  The properties of the starburst core of
M\,82, representative of those of the other selected regions, are
indicated by the horizontal bars.
\label{fig-paramvar}
}
\end{figure}

\clearpage

\begin{figure}[p]
\epsscale{2.0}
\plotone{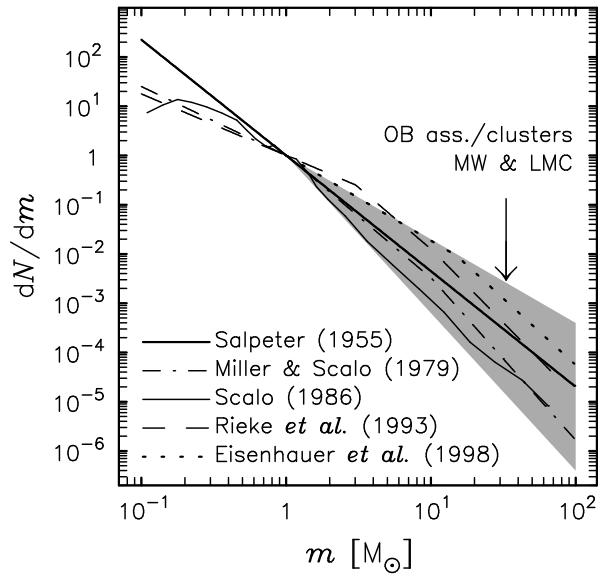}
\vspace{-4.0cm}
\caption{
Comparison of various IMFs, represented by different lines as labeled
in the plot.  The IMFs are normalized to unity at $\rm 1~M_{\odot}$.
The shaded area indicates the range of slopes determined in young clusters
and OB associations in the Milky Way and in the Large Magellanic Cloud
(\citealt{Hun97} and references therein; 
\citealt{Bra96, Eis98, Mas98, Sca98}).
\label{fig-IMF}
}
\end{figure}

\clearpage

\begin{figure}[p]
\epsscale{1.5}
\plotone{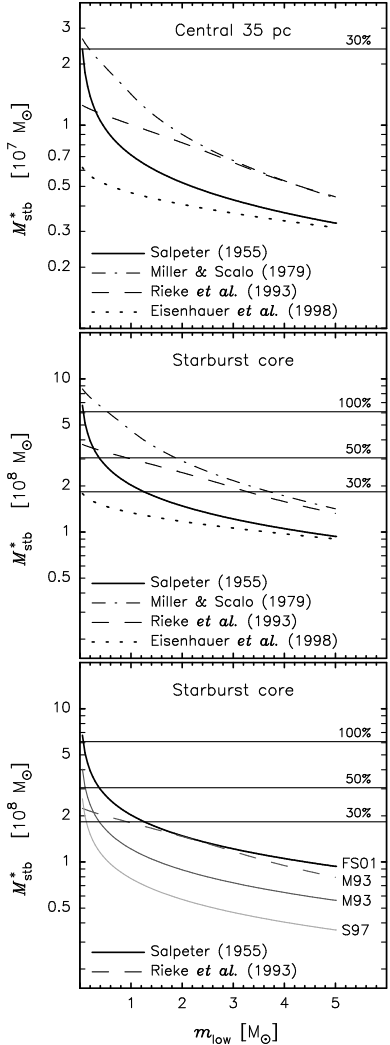}
\vspace{-1.0cm}
\caption{
Predicted mass involved in the starburst as a function of $m_{\rm low}$.
The curves show the burst masses for the nucleus and the starburst core
of M\,82 (central 35 and 500~pc) computed from the respective two-burst
models normalized such that the predicted luminosities ($L_{K}$ and
$L_{\rm Lyc}$) equal the observed luminosities for the derived burst
ages (see Tables~\ref{tab-modreg} and \ref{tab-modglob}).
Horizontal lines mark various fractions of the total stellar mass
determined for each region ($7.9 \times 10^{7}~{\rm M_{\odot}}$
and $6.1 \times 10^{8}~{\rm M_{\odot}}$ for the central 35~pc and
the starburst core, respectively).
The top and middle panels show, for each region, the effects of adopting
different IMFs, represented by the different curves as labeled in the plots.
The bottom panel illustrates results obtained for the starburst core
with a \citet{Sal55} IMF but assuming different values for the
$K$-band luminosities (solid curves; ``FS01'': this work, ``M93'':
\citealt{McL93}, ``S97'': \citealt{Sat97}; see \S~\ref{Sub-mlow}); 
the dashed curve show computations assuming the \citet{Rie93} IMF
and \citet{McL93} $L_{K}$.
\label{fig-mlow}
}
\end{figure}

\clearpage

\begin{figure}[p]
\epsscale{2.0}
\plotone{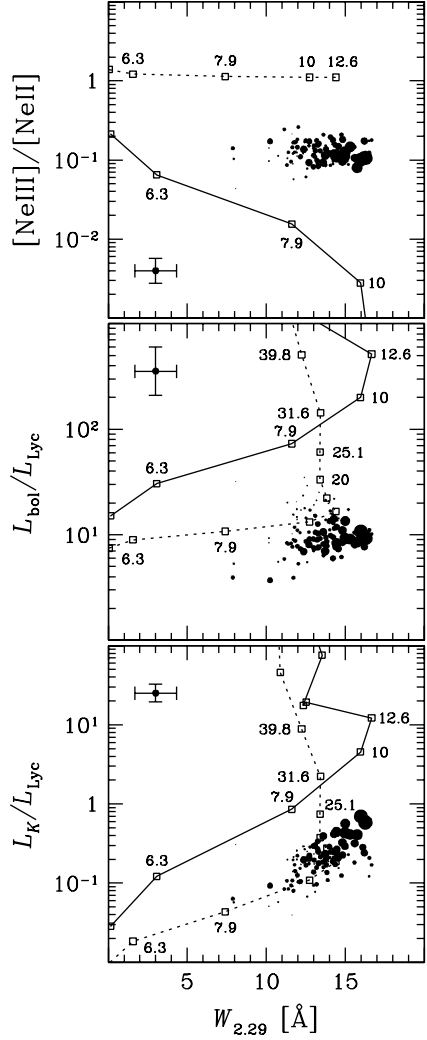}
\vspace{-1.0cm}
\caption{
Properties of individual $1^{\prime\prime} \times 1^{\prime\prime}$
pixels within the 3D field of view.
The diagrams are as in Figure~\ref{fig-modreg}, with model curves
computed for a single burst with $m_{\rm up} = 100~{\rm M_{\odot}}$,
and $t_{\rm sc} = 1$ and 5~Myr (solid and dashed lines, respectively);
open squares indicate burst ages separated by
$\Delta \log (t_{\rm b}\,[{\rm yr}]) = 0.1~{\rm dex}$,
with selected ages (in Myr) labeled for reference.
The size of the data points is proportional to the intrinsic stellar $L_{K}$.
Typical uncertainties are shown by the error bars in each plot.
\label{fig-modpix}
}
\end{figure}

\clearpage

\begin{figure}[p]
\epsscale{1.8}
\plotone{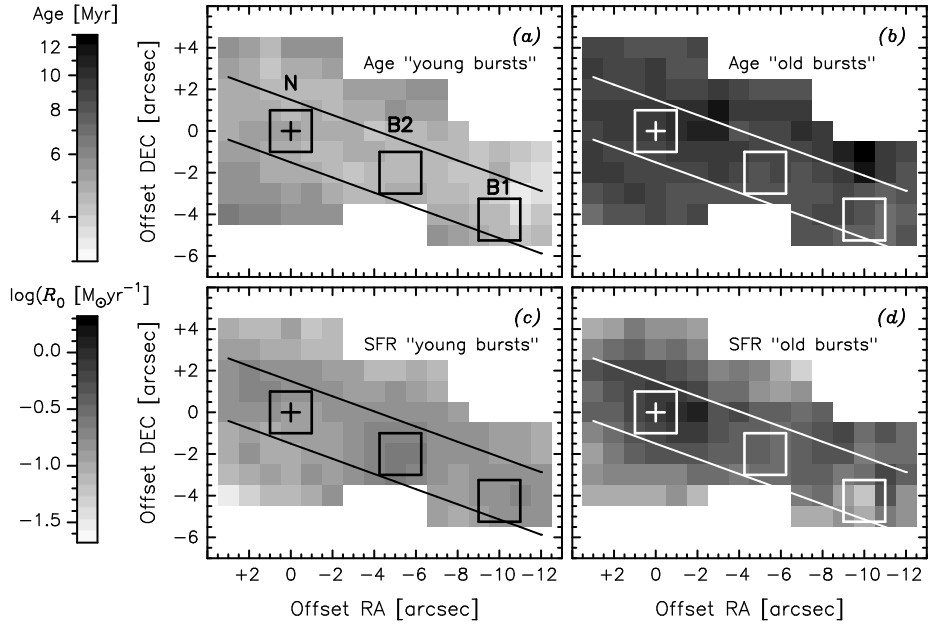}
\vspace{-3.5cm}
\caption{
Model results for individual pixels in the 3D field of view.
The top panels ({\em a\/} and {\em b}) show the derived ages for the
young and old bursts while the bottom panels ({\em c\/} and {\em d})
show the initial star formation rates ($R_{0}$).
The cross marks the position of the nucleus and square boxes indicate
selected regions modeled in \S~\ref{Sect-models_regions}: the
central 35~pc at the nucleus and regions B1 and B2.
The diagonal lines indicate the $3^{\prime\prime}$ --wide slit along
the galactic plane of M\,82 used to extract the profiles shown in 
Figure~\ref{fig-resprof}.
\label{fig-resmaps}
}
\end{figure}

\clearpage

\begin{figure}[p]
\epsscale{1.6}
\plotone{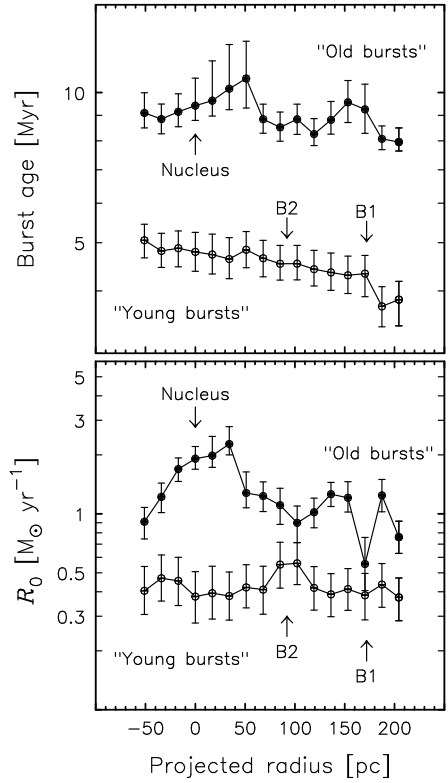}
\vspace{-2.0cm}
\caption{
Variations of the burst age and strength with projected radius from the
nucleus of M\,82.  The radial profiles are obtained from the results of
individual pixels in the $3^{\prime\prime}$ --wide slit along the galactic
plane shown in Figure~\ref{fig-resmaps}.  The results for the young and for
the old bursts are represented by open and filled circles, respectively.
The positions of the nucleus and of the regions B1 and B2 are indicated.
\label{fig-resprof}
}
\end{figure}

\clearpage

\begin{figure}[p]
\epsscale{1.6}
\plotone{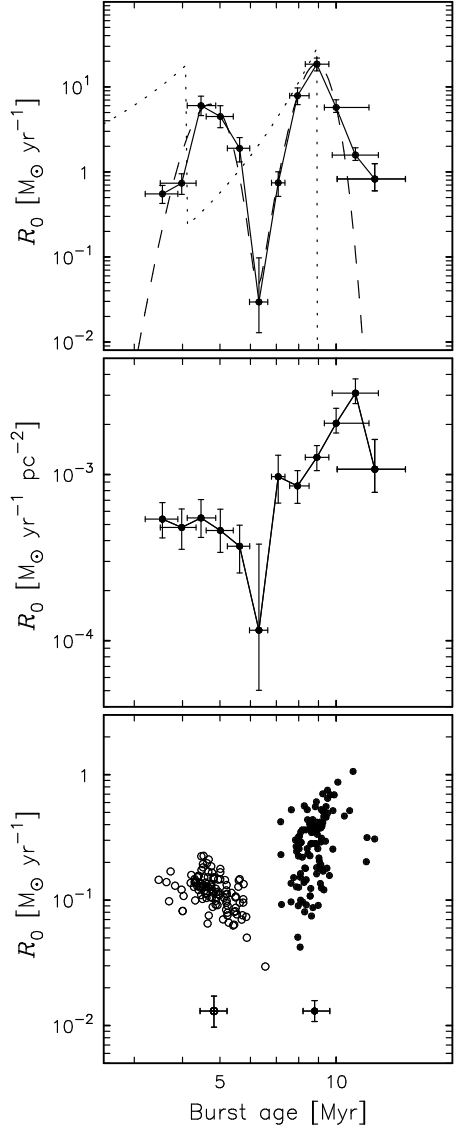}
\vspace{-0.8cm}
\caption{
Global star formation history within the 3D field of view in M\,82.
{\em Top panel\/}: integrated star formation rate derived from the
spatially detailed modeling and double Gaussian profile which provides
a good approximation thereof (solid line with filled dots and dashed
line, respectively; see \S~\ref{Sub-global}).  The best-fit two-burst
model for the global properties of the 3D field of view is plotted
for comparison (from Table~\ref{tab-modglob}; dotted line).
{\em Middle panel\/}: initial star formation rate surface density, 
obtained by dividing the $R_{0}$ versus burst age curve of the top panel
by the total area of the pixels contributing in each age bin.
{\em Bottom panel\/}: model results for the individual pixels, with open
and filled circles representing the young and old bursts, respectively.
Typical error bars are indicated at the bottom of the plot.
\label{fig-SFH}
}
\end{figure}

\clearpage

\begin{figure}[p]
\epsscale{1.8}
\plotone{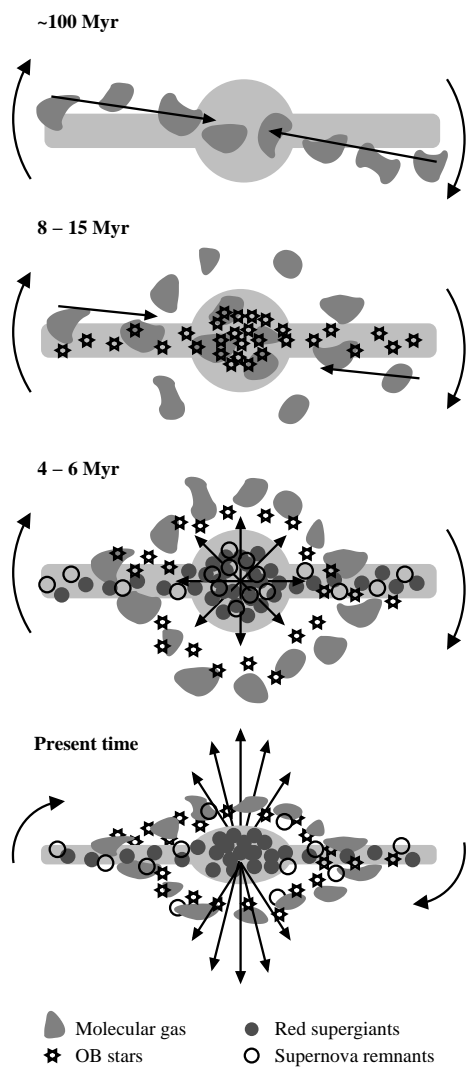}
\vspace{-2.0cm}
\caption{
Sketch of the succession of events related to starburst activity in
M\,82 depicted in \S~\ref{Sub-SFevol}.  
The galaxy is illustrated as viewed from above the plane except for 
the bottom panel, where it is inclined to show the starburst wind.
\label{fig-picture}
}
\end{figure}

\clearpage

\begin{figure}[p]
\epsscale{1.9}
\plotone{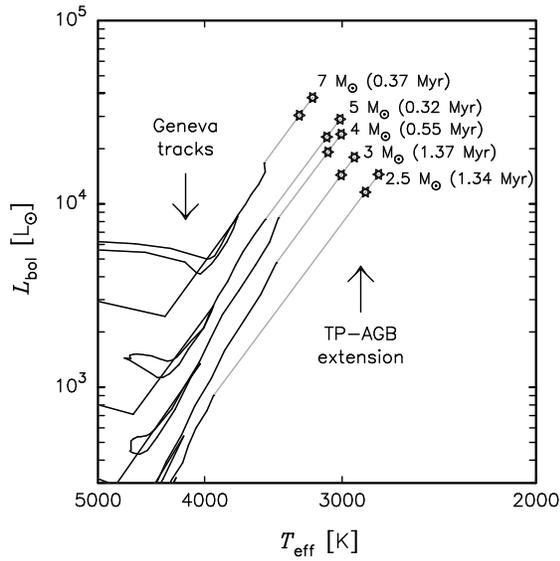}
\vspace{-4.0cm}
\caption{
Selected evolutionary tracks in the theoretical Hertzsprung-Russell diagram
for intermediate-mass stars.  Black lines represent the Geneva tracks.
Grey lines show our extensions to the TP-AGB phase computed from the
prescriptions of \citet{Bed88}; the star symbols correspond to the two
evolutionary points associated with the different pulsation modes
considered (appendix~\ref{App-STARS}).
Total TP-AGB lifetimes are given in parenthesis.
\label{fig-TPAGB}
}
\end{figure}

\clearpage

\begin{figure}[p]
\epsscale{2.0}
\plotone{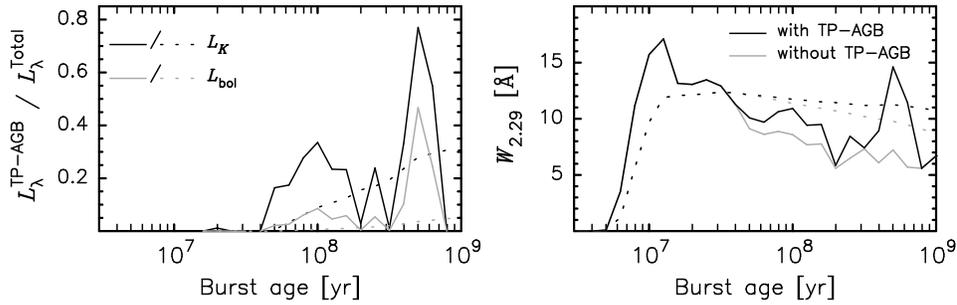}
\vspace{-7.0cm}
\caption{
Effects of TP-AGB stars in our starburst models.
The left panel shows the relative contribution of the TP-AGB phase
to the total $K$-band and bolometric luminosities (black and grey lines,
respectively).  The right panel compares the model predictions for the EW
of the CO bandhead at $\rm 2.29~\mu m$ accounting for TP-AGB stars (black
lines) and excluding TP-AGB stars (grey lines).  Computations are shown for
burst decay timescales of $t_{\rm sc} = 1~{\rm Myr}$ and 1~Gyr (solid and
dotted lines, respectively).  The effects of TP-AGB stars at ages
$\rm \sim 500~Myr$ for $t_{\rm sc} = 1~{\rm Myr}$ are likely exaggerated
due to the conventional synthesis technique employed in our
models (appendix~\ref{App-STARS}).
\label{fig-TPAGBmod}
}
\end{figure}

\clearpage

\begin{deluxetable}{lcccccc}
\tabletypesize{\small}
\tablecolumns{7}
\tablewidth{0pt}
\tablenum{1}
\tablecaption{Observational constraints for selected regions in M\,82
              \label{tab-Obs}}
\tablehead{
\colhead{Property} & \colhead{Units} & 
\colhead{Nucleus} & \colhead{B1} & \colhead{B2} & 
\colhead{3D field} & \colhead{Starburst core} 
}
\startdata
$L_{K}$ & $\rm 10^{8}\,L_{\odot}$ & $0.56 \pm 0.23$ & 
$0.074 \pm 0.024$ & $0.15 \pm 0.05$ & $2.7 \pm 0.8$ & $13 \pm 4$ \\
$L_{\rm Lyc}$ & $\rm 10^{8}\,L_{\odot}$ & $0.67 \pm 0.25$ & 
$1.5 \pm 0.4$ & $2.0 \pm 0.9$ & $27 \pm 13$ & $77 \pm 23$ \\
$L_{\rm bol}$ & $\rm 10^{8}\,L_{\odot}$ & $18 \pm 9$ & 
$8.9 \pm 3.1$ & $12 \pm 4$ & $220 \pm 90$ & $660 \pm 120$ \\
$L_{\rm bol}^{\rm OB}$ & $\rm 10^{8}\,L_{\odot}$ & $6.6 \pm 2.0$ & 
$7.4 \pm 2.2$ & $9.2 \pm 2.7$ & $170 \pm 60$ & $390 \pm 40$ \\
$M^{\star}$ & $\rm 10^{8}\,M_{\odot}$ & $0.79_{-0.21}^{+0.22}$ & 
... & ... & ... & $6.1_{-2.5}^{+2.7}$ \\
$\nu_{\rm SN}$\ \tablenotemark{a} & $\rm 10^{-2}\,yr^{-1}$ & 0.25 & 
0.25 & 0.58 & 13 & 6 \\
$W_{1.62}$ & \AA & $5.6 \pm 0.3$ & 
$3.4 \pm 0.3$ & $4.6 \pm 0.3$ & $4.8 \pm 0.3$ & $4.8 \pm 0.3$ \\
$W_{2.29}$ & \AA & $15.2 \pm 1.2$ & 
$11.2 \pm 1.4$ & $14.4 \pm 1.4$ & $14.4 \pm 1.4$ & $14.4 \pm 1.4$ \\
\hline
$L_{K}/L_{\rm Lyc}$ & ... & $0.84 \pm 0.46$ & 
$0.049 \pm 0.021$ & $0.075 \pm 0.042$ & $0.10 \pm 0.06$ & $0.17 \pm 0.07$ \\
$L_{\rm bol}/L_{\rm Lyc}$ & ... & $27 \pm 17$ & 
$5.9 \pm 2.6$ & $6.0 \pm 3.4$ & $8.1 \pm 5.1$ & $8.6 \pm 3.0$ \\
$L_{\rm bol}^{\rm OB}/L_{\rm Lyc}$ & ... & $9.9 \pm 4.7$ & 
$4.9 \pm 2.0$ & $4.6 \pm 2.5$ & $6.3 \pm 3.8$ & $5.1 \pm 1.6$ \\
$M^{\star}/L_{K}$ & $\rm M_{\odot}/L_{\odot}$ & $1.4 \pm 0.7$ & 
... & ... & ... & $0.47 \pm 0.25$ \\
$10^{12}\,\nu_{\rm SN}/L_{\rm bol}$\ \tablenotemark{a} & 
$\rm yr^{-1}/L_{\odot}$ & 1.4 & 2.8 & 4.8 & 5.9 & 0.91 \\
$\rm [Ne\,III]/[Ne\,II]$\ \tablenotemark{b} & ... &
0.13 & 0.16 & 0.12 & 0.16 & $0.16 \pm 0.04$ \\
\enddata
\tablenotetext{a}
{
Uncertainties of a factor of two are adopted for
$\nu_{\rm SN}$ and $10^{12}\,\nu_{\rm SN}/L_{\rm bol}$.
}
\tablenotetext{b}
{
Values for the nucleus, B1, B2, and the 3D field of view
are the inferred ``equivalent ratios'' as described in \S~\ref{Sect-Obs}.
The uncertainties are estimated to be a factor of two.
}
\end{deluxetable}

\clearpage

\begin{deluxetable}{lcl}
\tablecolumns{3}
\tablewidth{0pt}
\tablenum{2}
\tablecaption{Summary of model parameters    \label{tab-param}}
\tablehead{
\colhead{Parameter} & \colhead{Symbol} & \colhead{Value}
}
\startdata
\multicolumn{3}{c}{IMF: ${\rm d}N/{\rm d}m \propto m^{-\alpha}$} \\
Power-law index & $\alpha$ & $2.35$ (\citealt{Sal55}) \\
Upper mass cutoff & $m_{\rm up}$ & $25~{\rm M_{\odot}} - 100~{\rm M_{\odot}}$ \\
Lower mass cutoff & $m_{\rm low}$ & $0.1~{\rm M_{\odot}} - 5~{\rm M_{\odot}}$ \\
\hline
\multicolumn{3}{c}{Star formation history} \\
Burst timescale & $t_{\rm sc}$ & $1~{\rm Myr} - 1~{\rm Gyr}$ \\
Burst age & $t_{\rm b}$ & $1~{\rm Myr} - 100~{\rm Myr}$ \\
Burst strength\tablenotemark{a} & $R_{0}$ & \multicolumn{1}{c}{...} \\
\hline
\multicolumn{3}{c}{Stellar properties} \\
Metallicity & \multicolumn{1}{c}{...} & Solar \\
Mass-loss rate & \multicolumn{1}{c}{...} & Normal \\
\hline
\multicolumn{3}{c}{Nebular properties} \\
Electron density & $n_{\rm e}$ & $300~{\rm cm^{-3}}$ \\
Ionization parameter\tablenotemark{b} & $\log U_{\rm eff}$ & $-2.3~{\rm dex}$ \\
Inner radius\tablenotemark{b} & $R_{\rm eff}$ & 25~pc \\
Gas-phase abundances & \multicolumn{1}{c}{...} & Solar \\
Dust within nebulae & \multicolumn{1}{c}{...} & Neglected \\
\enddata
\tablenotetext{a}
{
Characterized by the initial star formation rate.
}
\tablenotetext{b}
{
Effective values representing the local nebular conditions in
the idealized thin gas shell geometry (see \S~\ref{Sub-CLOUDY}).
}
\end{deluxetable}

\clearpage

\begin{deluxetable}{lccccc}
\tabletypesize{\small}
\tablecolumns{6}
\tablewidth{0pt}
\tablenum{3}
\tablecaption{Starburst models for the nucleus and regions B1 and B2
              \label{tab-modreg}}
\tablehead{\multicolumn{6}{c}{{\bf Nucleus}}}
\startdata
\ Parameter & Units & 
\multicolumn{2}{c}{Young burst} & \multicolumn{2}{c}{Old burst} \\
\hline
\ $t_{\rm b}$ & $\rm Myr$ & 
\multicolumn{2}{c}{$4.7 \pm 0.5$} & \multicolumn{2}{c}{$9.6^{+0.6}_{-0.2}$} \\
\ $R_{0}$ & $\rm M_{\odot}\,yr^{-1}$ & 
\multicolumn{2}{c}{$0.52^{+0.30}_{-0.25}$} & 
\multicolumn{2}{c}{$6.6 \pm 1.5$} \\
\hline
\ Property & Units & Observed & Young burst & Old burst & Total \\
\hline
\ $L_{K}$ & $\rm 10^{8}~L_{\odot}$ & $0.56 \pm 0.23$ & 
0.011 & 0.55 & 0.56 \\
\ $L_{\rm Lyc}$ & $\rm 10^{8}~L_{\odot}$ & $0.67 \pm 0.25$ & 
0.48 & 0.16 & 0.64 \\
\ $L_{\rm bol}$ & $\rm 10^{8}~L_{\odot}$ & $18 \pm 9$ & 
6.8 & 27 & 34 \\
\ $M^{\star}$ & $\rm 10^{8}~M_{\odot}$ & $0.79^{+0.22}_{-0.21}$ & 
0.0052 & 0.066 & 0.071 \\
\ $W_{1.62}$ & \AA & $5.6 \pm 0.3$ & 
0.0 & 5.6 & 5.5 \\
\ $W_{2.29}$ & \AA & $15.2 \pm 1.2$ & 
0.0 & 15.2 & 14.9 \\
\ $\nu_{\rm SN}$\,\tablenotemark{a} & $\rm 10^{-2}\,yr^{-1}$ & 0.25 & 
0.038 & 0.45 & 0.49 \\ 
\ [\ion{Ne}{3}]/[\ion{Ne}{2}]\,\tablenotemark{a} & ... & 0.13 &
0.25 & 0.0038 & 0.21 \\ 
\hline\hline
\multicolumn{6}{c}{{\bf B1}} \\
\hline
\ Parameter & Units & 
\multicolumn{2}{c}{Young burst} & \multicolumn{2}{c}{Old burst} \\
\hline
\ $t_{\rm b}$ & $\rm Myr$ & 
\multicolumn{2}{c}{$3.9^{+0.4}_{-0.6}$} & 
\multicolumn{2}{c}{$7.8^{+0.2}_{-0.1}$} \\
\ $R_{0}$ & $\rm M_{\odot}\,yr^{-1}$ & 
\multicolumn{2}{c}{$0.84 \pm 0.26$} & 
\multicolumn{2}{c}{$0.96^{+0.27}_{-0.30}$} \\
\hline
\ Property & Units & Observed & Young burst & Old burst & Total \\
\hline
\ $L_{K}$ & $\rm 10^{8}~L_{\odot}$ & $0.074 \pm 0.024$ & 
0.013 & 0.061 & 0.074 \\
\ $L_{\rm Lyc}$ & $\rm 10^{8}~L_{\odot}$ & $1.5 \pm 0.4$ & 
1.4 & 0.083 & 1.5 \\
\ $L_{\rm bol}$ & $\rm 10^{8}~L_{\odot}$ & $8.9 \pm 3.1$ & 
14 & 5.7 & 20 \\
\ $M^{\star}$ & $\rm 10^{8}~M_{\odot}$ & ... & 
0.0082 & 0.0096 & 0.018 \\
\ $W_{1.62}$ & \AA & $3.4 \pm 0.3$ & 
0.0 & 3.5 & 2.9 \\
\ $W_{2.29}$ & \AA & $11.2 \pm 1.4$ & 
0.0 & 10.9 & 9.0 \\
\ $\nu_{\rm SN}$\,\tablenotemark{a} & $\rm 10^{-2}\,yr^{-1}$ & 0.25 & 
0.040 & 0.078 & 0.12 \\
\ [\ion{Ne}{3}]/[\ion{Ne}{2}]\,\tablenotemark{a} & ... & 0.16 &
0.45 & 0.017 & 0.42 \\ 
\hline\hline
\multicolumn{6}{c}{{\bf B2}} \\
\hline
\ Parameter & Units & 
\multicolumn{2}{c}{Young burst} & \multicolumn{2}{c}{Old burst} \\
\hline
\ $t_{\rm b}$ & $\rm Myr$ & 
\multicolumn{2}{c}{$4.0^{+0.4}_{-0.5}$} & \multicolumn{2}{c}{$8.9 \pm 0.2$} \\
\ $R_{0}$ & $\rm M_{\odot}\,yr^{-1}$ & 
\multicolumn{2}{c}{$1.2^{+0.5}_{-0.4}$} & 
\multicolumn{2}{c}{$1.7^{+0.4}_{-0.5}$} \\
\hline
\ Property & Units & Observed & Young burst & Old burst & Total \\
\hline
\ $L_{K}$ & $\rm 10^{8}~L_{\odot}$ & $0.15 \pm 0.05$ & 
0.019 & 0.13 & 0.15 \\
\ $L_{\rm Lyc}$ & $\rm 10^{8}~L_{\odot}$ & $2.0 \pm 0.9$ & 
1.8 & 0.067 & 1.9 \\
\ $L_{\rm bol}$ & $\rm 10^{8}~L_{\odot}$ & $12 \pm 4$ & 
19 & 8.0 & 27 \\
\ $M^{\star}$ & $\rm 10^{8}~M_{\odot}$ & ... & 
0.012 & 0.017 & 0.029 \\
\ $W_{1.62}$ & \AA & $4.6 \pm 0.3$ & 
0.0 & 4.9 & 4.2 \\
\ $W_{2.29}$ & \AA & $14.4 \pm 1.4$ & 
0.0 & 13.8 & 12.0 \\
\ $\nu_{\rm SN}$\,\tablenotemark{a} & $\rm 10^{-2}\,yr^{-1}$ & 0.58 & 
0.062 & 0.13 & 0.19 \\ 
\ [\ion{Ne}{3}]/[\ion{Ne}{2}]\,\tablenotemark{a} & ... & 0.12 &
0.41 & 0.0067 & 0.40 \\
\enddata
\tablenotetext{a}
{
The adopted uncertainties for the constraints on the neon ratio and 
\snrate\ are a factor of two.
}
\end{deluxetable}

\clearpage

\begin{deluxetable}{lccccc}
\tabletypesize{\small}
\tablecolumns{6}
\tablewidth{0pt}
\tablenum{4}
\tablecaption{Starburst models for the 3D field of view and the starburst core
              \label{tab-modglob}}
\tablehead{\multicolumn{6}{c}{{\bf 3D field of view}}}
\startdata
\ Parameter & Units & 
\multicolumn{2}{c}{Young burst} & \multicolumn{2}{c}{Old burst} \\
\hline
\ $t_{\rm b}$ & $\rm Myr$ & 
\multicolumn{2}{c}{$4.1^{+0.5}_{-0.7}$} & \multicolumn{2}{c}{$9.0 \pm 0.2$} \\ 
\ $R_{0}$ & $\rm M_{\odot}\,yr^{-1}$ & 
\multicolumn{2}{c}{$18^{+9}_{-8}$} & \multicolumn{2}{c}{$31^{+7}_{-8}$} \\ 
\hline
\ Property & Units & Observed & Young burst & Old burst & Total \\
\hline
\ $L_{K}$ & $\rm 10^{8}~L_{\odot}$ & $2.7 \pm 0.8$ & 
0.30 & 2.4 & 2.7 \\
\ $L_{\rm Lyc}$ & $\rm 10^{8}~L_{\odot}$ & $27 \pm 13$ & 
25 & 1.1 & 26 \\
\ $L_{\rm bol}$ & $\rm 10^{8}~L_{\odot}$ & $220 \pm 90$ & 
280 & 140 & 420 \\
\ $M^{\star}$ & $\rm 10^{8}~M_{\odot}$ & ... & 
0.18 & 0.31 & 0.49 \\
\ $W_{1.62}$ & \AA & $4.8 \pm 0.3$ & 
0.0 & 5.0 & 4.4 \\
\ $W_{2.29}$ & \AA & $14.4 \pm 1.4$ & 
0.0 & 14.0 & 12.4 \\
\ $\nu_{\rm SN}$\,\tablenotemark{a} & $\rm 10^{-2}\,yr^{-1}$ & 13 & 
0.99 & 2.3 & 3.3 \\
\ [\ion{Ne}{3}]/[\ion{Ne}{2}]\,\tablenotemark{a} & ... & 0.16 &
0.38 & 0.0061 & 0.37 \\ 
\hline\hline
\multicolumn{6}{c}{{\bf Starburst core}} \\
\hline
\ Parameter & Units & 
\multicolumn{2}{c}{Young burst} & \multicolumn{2}{c}{Old burst} \\
\hline
\ $t_{\rm b}$ & $\rm Myr$ & 
\multicolumn{2}{c}{$3.9^{+0.2}_{-0.3}$} & \multicolumn{2}{c}{$9.0 \pm 0.2$} \\
\ $R_{0}$ & $\rm M_{\odot}\,yr^{-1}$ & 
\multicolumn{2}{c}{$43 \pm 11$} & \multicolumn{2}{c}{$160 \pm 30$} \\
\hline
\ Property & Units & Observed & Young burst & Old burst & Total \\
\hline
\ $L_{K}$ & $\rm 10^{8}~L_{\odot}$ & $13 \pm 4$ & 
0.66 & 12 & 13 \\
\ $L_{\rm Lyc}$ & $\rm 10^{8}~L_{\odot}$ & $77 \pm 23$ & 
69 & 5.8 & 75 \\
\ $L_{\rm bol}$ & $\rm 10^{8}~L_{\odot}$ & $660 \pm 120$ & 
710 & 740 & 1500 \\
\ $M^{\star}$ & $\rm 10^{8}~M_{\odot}$ & $6.1^{+2.7}_{-2.5}$ & 
0.42 & 1.6 & 2.0 \\
\ $W_{1.62}$ & \AA & $4.8 \pm 0.3$ & 
0.0 & 5.0 & 4.7 \\
\ $W_{2.29}$ & \AA & $14.4 \pm 1.4$ & 
0.0 & 14.0 & 13.3 \\
\ $\nu_{\rm SN}$\,\tablenotemark{a} & $\rm 10^{-2}\,yr^{-1}$ & 6 & 
2.1 & 12 & 14 \\ 
\ [\ion{Ne}{3}]/[\ion{Ne}{2}] & ... & $0.16 \pm 0.04$ &
0.45 & 0.0061 & 0.42 \\ 
\enddata
\tablenotetext{a}
{
The adopted uncertainties for the constraints on the neon ratio for
the 3D field of view and on \snrate\ for both regions are a factor of two.
}
\end{deluxetable}

\clearpage

\begin{deluxetable}{lccccc}
\tablecolumns{6}
\tablewidth{0pt}
\tablenum{5}
\tablecaption{Parameters for various IMFs\tablenotemark{a}   \label{tab-IMF}}
\tablehead{
\colhead{Source} & \colhead{$\alpha_{1}$} & \colhead{$m_{1}$} & 
\colhead{$\alpha_{2}$} & \colhead{$m_{2}$} & \colhead{$\alpha_{3}$} \\
 &  & \colhead{($\rm M_{\odot}$)} &  & \colhead{($\rm M_{\odot}$)} & 
}
\startdata
\citealt{Sal55} & 2.35 & ... & ... & ... & ... \\
\citealt{MS79} & 1.4 & 1 & 2.5 & 10 & 3.3 \\
\citealt{Rie93} & 1.25 & 3 & 2.5 & 10 & 3.0 \\
\citealt{Eis98} & 1.73 & 15 & 2.7 & ... & ... \\
\enddata
\tablenotetext{a}
{
The IMFs are represented by a power-law ${\rm d}N/{\rm d}m \propto m^{-\alpha}$.
The values of $\alpha_{i}$ give the power-law indices for different mass ranges
and the $m_{i}$ correspond to the inflection points for broken power-laws.
}
\end{deluxetable}


\clearpage


\begin{thebibliography}{}

\bibitem[Achtermann \& Lacy(1995)]{Ach95} Achtermann, J. M. \& Lacy, J. H.  
	1995, \apj, 439, 163
\bibitem[Adelberger \& Steidel(2000)]{Ade00} 
        Adelberger, K. L., \& Steidel, C. C. 2000, \apj, 544, 218
\bibitem[Alonso-Herrero \etal(2001)]{Alo01} Alonso-Herrero, A., 
        Engelbracht, C. W., Rieke, M. J., Rieke, G. H., \& Quillen, A. C.
        2001, \apj, 546, 952
\bibitem[Alonso-Herrero \etal(2003)]{Alo03} Alonso-Herrero, A., 
        Rieke, G. H., Rieke, M. J., \& Kelly, D. M.
        2003, \aj, 125, 1210
\bibitem[Athanassoula(1992)]{Ath92} Athanassoula, E. 1992, \mnras, 259, 345
\bibitem[Augarde \& Lequeux(1985)]{Aug85} Augarde, R. \& Lequeux, J. 
        1985, \aap, 147, 273
\bibitem[Bedijn(1988)]{Bed88} Bedijn, P. J. 1988, \aap, 205, 105
\bibitem[Bernl\"ohr(1992)]{Ber92} Bernl\"ohr, K. 1992, \aap, 263, 54
\bibitem[Blum \etal(1996)Blum, Sellgren, \& DePoy]{Blu96} 
        Blum, R. D., Sellgren, K., \& DePoy, D. L. 1996, \aj, 112, 1988
\bibitem[B\"oker \etal(1997)B\"oker, F\"orster Schreiber, \& Genzel]{Bok97}
        B\"oker, T., F\"orster Schreiber, N. M., \& Genzel, R.
	1997, \aj, 114, 1883
\bibitem[Brandl \etal(1996)]{Bra96} Brandl, B., \etal 1996, \apj, 466, 254
\bibitem[Brandl \etal(1999)]{Bra99} Brandl, B., Brandner, W., Eisenhauer, F.,
        Moffat, A. F. J., Palla, F., \& Zinnecker, H. 1999, \aap, 352, L69
\bibitem[Bregman \etal(1995)Bregman, Schulman, \& Tomisaka]{Bre95} 
        Bregman, J. N., Schulman, E., \& Tomisaka, K. 1995, \apj, 439, 155
\bibitem[Bruzual \& Charlot(1993)]{BC93} Bruzual A., G. \& Charlot, S.
        1993, \apj, 405, 538
\bibitem[Carlstrom \& Kronberg(1991)]{Car91} Carlstrom, J. E. \& Kronberg, P. P.
	1991, \apj, 366, 422
\bibitem[Cervi\~{n}o \& Mas-Hesse(1994)]{Cer94} Cervi\~{n}o, M. 
        \& Mas-Hesse, J. M. 1994, \aap, 284, 749
\bibitem[Chapman \etal(2003)]{Cha03} Chapman, S. C., Blain, A. W.,
         Ivison, R. J., \& Smail, I. R. 2003, \nat, 422, 695
\bibitem[Charlot \& Bruzual(1991)]{CB91} Charlot, S. \& Bruzual A., G. 
        1991, \apj, 367, 126
\bibitem[Chevalier \& Clegg(1985)]{Che85} Chevalier, R. A. \& Clegg, A. W. 
        1985, \nat, 317, 44
\bibitem[Colbert \etal(1999)]{Col99} Colbert, J. W., \etal 1999, \apj, 511, 721
\bibitem[Combes \& G\'erin(1985)]{Com85} Combes, F. \& G\'erin, M. 
        1985, \aap, 150, 327 
\bibitem[Davidge \etal(1997)]{Dav97} Davidge, T. J., Rigaut, F., Doyon, R.,
        \& Crampton, D. 1997, \aj, 113, 2094
\bibitem[de Graauw \etal(1996)]{deG96} de Graauw, T., \etal 1996, \aap, 315, L49
\bibitem[de Grijs \etal(2000)]{Gri00} de Grijs, R., O'Connell, R. W.,
        Becker, G. D., Chevalier, R. A., \& Gallagher, J. S., III.
        2000, \aj, 119, 681
\bibitem[de Grijs \etal(2001)de Grijs, O'Connell, \& Gallagher]{Gri01} 
        de Grijs, R., O'Connell, R. W., \& Gallagher, J. S., III.
        2001, \aj, 121, 768 
\bibitem[Devereux \etal(1987)Devereux, Becklin, \& Scoville]{Dev87} 
        Devereux, N. A., Becklin, E. E., \& Scoville, N. Z. 1987, \apj, 312, 529
\bibitem[Doane \& Mathews(1993)]{Doa93} Doane, J. S. \& Mathews, W. G.  
        1993, \apj, 419, 573
\bibitem[Eisenhauer \etal(1998)]{Eis98} Eisenhauer, F., Quirrenbach, A., 
        Zinnecker, H., \& Genzel, R. 1998, \apj, 498, 278
\bibitem[Engelbracht \etal(1998)]{Eng98} Engelbracht, C. W., Rieke, M. J.,
        Rieke, G. H., Kelly, D. M., \& Achtermann, J. M. 1998, \apj, 505, 639
\bibitem[Engelbracht \etal(1996)]{Eng96} Engelbracht, C. W., Rieke, M. J.,
        Rieke, G. H., \& Latter, W. B. 1996, \apj, 467, 227
\bibitem[Ferland(1996)]{Ferl96} Ferland, G. J. 1996, Hazy, a Brief Introduction
        to Cloudy, University of Kentucky Department of Physics and 
        Astronomy Internal Report
\bibitem[Fioc \& Rocca-Volmerange(1997)]{Fio97} Fioc, M. \& Rocca-Volmerange, B.
        1997, \aap, 326, 950
\bibitem[F\"orster Schreiber(2000)]{FS00} F\"orster Schreiber, N. M.
        2000, \aj, 120, 2089
\bibitem[F\"orster Schreiber \etal(2001)]{FS01} F\"orster Schreiber, N. M.,
        Genzel, R., Lutz, D., Kunze, D., \& Sternberg, A.
        2001, \apj, 552, 544
\bibitem[Freedman(1992)]{Fre92} Freedman, W. L. 1992, \aj, 104, 1349
\bibitem[Freedman \& Madore(1988)]{Fre88} Freedman, W. L. \& Madore, B. F. 
        1988, \apj, 332, L63
\bibitem[Gaffney \etal(1993)Gaffney, Lester, \& Telesco]{Gaf93}
        Gaffney, N. I., Lester, D. F., \& Telesco, C. M. 1993, \apj, 407, L57
\bibitem[Gallagher \& Smith(1999)]{Gal99} Gallagher, J. S., III \& Smith, L. J.
        1999, \mnras, 304, 540 
\bibitem[Giavalisco(2002)]{Gia02} Giavalisco, M. 2002, \araa, 40, 579
\bibitem[Giveon \etal(2002)]{Giv02} Giveon, U., Sternberg, A., Lutz, D.,
        Feuchtgruber, H., \& Pauldrach, A. W. A.
	2002, \apj, 566, 880
\bibitem[Goldader \etal(1997)]{Gol97} Goldader, J. D., Joseph, R. D.,
        Doyon, R., \& Sanders, D. B. 1997, \apj, 474, 104
\bibitem[G\"otz \etal(1990)]{Got90} G\"otz, M., McKeith, C. D., Downes, D.,
        \& Greve, A. 1990, \aap, 240, 52
\bibitem[Gottesman \& Weliachew(1977)]{Got77} Gottesman, S. T. \& Weliachew, L.
        1977, \apj, 211, 47
\bibitem[Greenhouse \etal(1997)]{Gre97} Greenhouse, M. A., \etal
        1997, \apj, 476, 105
\bibitem[Grevesse \& Anders(1989)]{Gre89} Grevesse, N, \& Anders, E.
        1989, in AIP Conf. Proc. 183, Cosmic Abundances of Matter,
	ed. C. J. Waddington (New York: AIP), 1
\bibitem[Grevesse \& Noels(1993)]{Gre93} Grevesse, N, \& Noels, A.
        1989, in Origin and Evolution of the Elements,
	ed. N. Prantzos, E. Vangioni-Flam, \& M. Casse
	(Cambridge: Cambridge Univ. Press), 1
\bibitem[Heckman(1998)]{Hec98} Heckman, T. M. 1998, in ASP Conf. Ser. 148,
        Origins, ed. C. E. Woodward, J. M. Shull, \& H. A. Thronson Jr.
        (San Francisco: ASP), 127
\bibitem[Heckman \etal(1990)Heckman, Armus, \& Miley]{Hec90} 
        Heckman, T. M., Armus, L., \& Miley, G. K. 1990, \apjs, 74, 833
\bibitem[Heller \& Shlosman(1994)]{Hel94} Heller, C. H. \& Shlosman, I. 
        1994, \apj, 424, 84
\bibitem[Hernquist(1989)]{Her89} Hernquist, L. 1989, \nat, 340, 687
\bibitem[Ho \& Fillipenko(1996)]{Ho96} Ho, L. C. \& Fillipenko, A. V. 
        1996, \apj, 472, 600
\bibitem[Hunt \etal(1999)]{Hun99} Hunt, L. K., Malkan, M. A., Moriondo, G.,
        \& Salvati, M. 1999, \apj, 510, 637
\bibitem[Hunter \etal(1997)]{Hun97} Hunter, D. A., Light, R. M., 
        Holtzman, J. A., Lynds, R., O'Neil, E. J., Jr., \& Grillmair, C. J.
        1997, \apj, 478, 124 
\bibitem[Kennicutt(1992)]{Ken92} Kennicutt, R. C., Jr. 1992, \apj, 388, 310
\bibitem[Kennicutt(1998)]{Ken98} Kennicutt, R. C., Jr. 1998, \apj, 498, 541
\bibitem[Kessler \etal(1996)]{Kes96} Kessler, M. F., \etal 1996, \aap, 315, L27
\bibitem[Knapen \etal(1995)]{Kna95} Knapen, J. H., Beckman, J. E., 
        Heller, C. H., Shlosman, I., \& de Jong, R. S. 1995, \apj, 454, 623
\bibitem[Koornneef(1983)]{Koo83} Koornneef, J. 1983, \aap, 128, 84
\bibitem[Kronberg \etal(1985)Kronberg, Biermann, \& Schwab]{Kro85}
	Kronberg, P. P., Biermann, P., \& Schwab, F. R. 1985, \apj, 291, 693
\bibitem[Kroupa(2001)]{Kro01} Kroupa, P. 2001, \mnras, 322, 231
\bibitem[Kurucz(1992)]{Kur92} Kurucz, R. L. 
        1992, Rev. Mex. Astron. Astrof., 23, 181
\bibitem[Lan\c{c}on \etal(1999)]{Lan99} Lan\c{c}on, A., Mouhcine, M., 
        Fioc, M., \& Silva, D. 1999, \aap, 344, L21
\bibitem[Lan\c{c}on \& Rocca-Volmerange(1994)]{LRV94} Lan\c{c}on, A. 
        \& Rocca-Volmerange, B. 1994, Ap{\&}SS, 217, 271
\bibitem[Lan\c{c}on \& Wood(2000)]{Lan00} Lan\c{c}on, A., \& Wood, P. R.
        2000, \aaps, 146, 217
\bibitem[Larkin \etal(1994)]{Lar94} Larkin, J. E., Graham, J. R., Matthews, K.,
	Soifer, B. T., Beckwith, S., Herbst, T. M., \& Quillen, A. C.  
	1994, \apj, 420, 159
\bibitem[Lee(1996)]{Lee96} Lee, M. G. 1996, \aj, 112, 1438
\bibitem[Lehnert \etal(1999)Lehnert, Heckman, \& Weaver]{Leh99} 
        Lehnert, M. D., Heckman, T. M., \& Weaver, K. A. 1999, \apj, 523, 575
\bibitem[Leitherer \etal(1999)]{Lei99} Leitherer, C., \etal 
        1999, \apjs, 123, 3  
\bibitem[Lo \etal(1987)]{Lo87} Lo, K. Y., Cheung, K. W., Masson, C. R., 
	Phillips, T. G., Scott, S. L., \& Woody, D. P.  
	1987, \apj, 312, 574
\bibitem[Maraston(1998)]{Maras98} Maraston, C. 1998, \mnras, 300, 872
\bibitem[Marcum \& O'Connell(1996)]{Mar96} Marcum, P., \& O'Connell, R. W.
        1996, in ASP Conf. Ser. 98, From Stars to Galaxies:
	The Impact of Stellar Physics on Galaxy Evolution, 
	ed. C. Leitherer, U. Fritze-von Alvensleben, \& J. Huchra
	(San Francisco: ASP), 419
\bibitem[Mart\'{\i}nez-Delgado \& Aparicio(1998)]{Mar98} 
        Mart\'{\i}nez-Delgado, D. \& Aparicio, A. 1998, \aj, 115, 1462
\bibitem[Massey \& Hunter(1998)]{Mas98} Massey, P. \& Hunter, D. A. 
        1998, \apj, 493, 180
\bibitem[McCrady \etal(2003)McCrady, Gilbert, \& Graham]{MCr03}
        McCrady, N., Gilbert, A. M., \& Graham, J. R.
        2003, \apj, in press (astro-ph/0306373)
\bibitem[McKeith \etal(1993)]{McK93} McKeith, C. D., Castles, J., 
        Greve, A., \& Downes, D. 1993, \aap, 272, 98
\bibitem[McLean \& Liu(1996)]{McL96} McLean, I. S. \& Liu, T. 
        1996, \apj, 456, 499
\bibitem[McLeod \etal(1993)]{McL93} McLeod, K. K., Rieke, G. H., Rieke, M. J.,
	\& Kelly, D. M. 1993, \apj, 412, 111  
\bibitem[Mendoza \& Johnson(1965)]{Men65} Mendoza, E. E., V, \& Johnson, H. L. 
        1965, \apj, 141, 161
\bibitem[Mengel \etal(2002)]{Men02} Mengel, S., Lehnert, M. D., Thatte, N.,
        \& Genzel, R. 2002, \aap, 383, 137
\bibitem[Meurer \etal(1995)]{Meu95} Meurer, G. R., Heckman, T. M., 
        Leitherer, C., Kinney, A., Robert, C., \& Garnett, D. R.
        1995, \aj, 110, 2665
\bibitem[Meynet \etal(1994)]{Mey94} Meynet, G., Maeder, A., Schaller, G., 
        Schaerer, D., \& Charbonnel, C. 1994, \aaps, 103, 97
\bibitem[Mihos \& Hernquist(1994)]{Mih94} Mihos, J. C. \& Hernquist, L. 
        1994, \apj, 431, L9
\bibitem[Mihos \& Hernquist(1996)]{Mih96} Mihos, J. C. \& Hernquist, L. 
        1996, \apj, 464, 641
\bibitem[Miller \& Scalo(1979)]{MS79} Miller, G. E. \& Scalo, J. M. 
        1979, \apjs, 41, 513
\bibitem[Mouhcine \& Lan\c{c}on(2002)]{Mou02} Mouhcine, M., \& Lan\c{c}on, A.
        2002, \aap, 393, 149
\bibitem[Muxlow \etal(1994)]{Mux94} Muxlow, T. W. B., Pedlar, A., 
        Wilkinson, P. N., Axon, D. J., Sanders, E. M., \& de Bruyn, A. G.
	1994, \mnras, 266, 455
\bibitem[Nakagawa \etal(1989)]{Nak89} Nakagawa, T., Nagata, T., Geballe, T. R., 
        Okuda, H., Shibai, H., \& Matsuhara, H. 1989, \apj, 340, 729 
\bibitem[Neininger \etal(1998)]{Nei98} Neininger, N., Gu\'elin, M., Klein, U., 
        Garc\'{\i}a-Burillo, S., \& Wielebinski, R. 1998, \aap, 339, 737
\bibitem[Noguchi(1987)]{Nog87} Noguchi, M. 1987, \mnras, 228, 635
\bibitem[Noguchi(1988)]{Nog88} Noguchi, M. 1988, \aap, 203, 259
\bibitem[O'Connell \& Mangano(1978)]{OCo78} O'Connell, R. W. \& Mangano, J. J.  
        1978, \apj, 221, 62
\bibitem[O'Connell \etal(1995)]{OCo95} O'Connell, R. W., Gallagher, J. S., III,
        Hunter, D., \& Colley, W. N. 1995, \apj, 446, L1
\bibitem[Oliva \etal(1995)]{OOKM95} Oliva, E., Origlia, L., Kotilainen, J. K.,
	\& Moorwood, A. F. M. 1995, \aap, 301, 55
\bibitem[Olofsson(1989)]{Olo89} Olofsson, K. 1989, \aaps, 80, 317
\bibitem[Origlia \etal(1999)]{Ori99} Origlia, L., Goldader, J. D.,
        Leitherer, C., Schaerer, D., \& Oliva, E. 1999, \apj, 514, 96
\bibitem[Origlia \& Oliva(2000)]{Ori00} Origlia, L. \& Oliva, E. 
        2000, \aap, 357, 61
\bibitem[Papovich \etal(2001)Papovich, Dickinson, \& Ferguson]{Pap01}
         Papovich, C., Dickinson, M., \& Ferguson, H. C. 2001, \apj, 559, 620
\bibitem[Pauldrach \etal(2001)Pauldrach, Hoffmann, \& Lennon]{Pau01}
        Pauldrach, A. W. A., Hoffmann, T. L., \& Lennon, M. 2001,
        \aap, 375, 161  
\bibitem[Pauldrach \etal(1998)]{Pau98} Pauldrach, A. W. A., Lennon, M., 
        Hoffmann, T. L., Sellmaier, F., Kudritzki, R.-P., \& Puls, J.
        1998, in ASP Conf. Ser. 131, Properties of Hot Luminous Stars,
        ed. I. Howarth (San Francisco: ASP), 258
\bibitem[Prestwich \etal(1994)Prestwich, Joseph, \& Wright]{Pre94} 
        Prestwich, A. H., Joseph, R. D., \& Wright, G.  S. 1994, \apj, 422, 73
\bibitem[Puxley \etal(1989)]{Pux89} Puxley, P. J., Brand, P. W. J. L., 
        Moore, T. J. T., Mountain, C. M., Nakai, N., \& Yamashita, T.
	1989, \apj, 345, 163
\bibitem[Puxley \etal(1997)]{Pux97} Puxley, P. J., Doyon, R., \& Ward, M. J.
        1997, \apj, 476, 120
\bibitem[Rana(1987)]{Ran87} Rana, N. C. 1987, \aap, 184, 104
\bibitem[Renzini \& Buzzoni(1986)]{Ren86} Renzini, A., \& Buzzoni, A.
        1986, in Spectral Evolution of Galaxies, ed. C. Chiosi \& A. Renzini
	(Dordrecht: Reidel), 195
\bibitem[Rieke \etal(1980)]{Rie80} Rieke, G. H., Lebofsky, M. J., 
        Thompson, R. I., Low, F. J., \& Tokunaga, A. T. 1980, \apj, 238, 24
\bibitem[Rieke \etal(1993)]{Rie93} Rieke, G. H., Loken, K., Rieke, M. J.,
	\& Tamblyn, P. 1993, \apj, 412, 99
\bibitem[Salpeter(1955)]{Sal55} Salpeter, E. E. 1955, \apj, 121, 161
\bibitem[Satyapal \etal(1997)]{Sat97} Satyapal, S., Watson, D. M., 
        Pipher, J. L., Forrest, W. J., Greenhouse, M. A., Smith, H.A.,
	Fisher, J., \& Woodward, C. E. 1997, \apj, 483, 148
\bibitem[Scalo(1986)]{Sca86} Scalo, J. M. 1986, Fund. Cosmic Phys., 11, 1
\bibitem[Scalo(1998)]{Sca98} Scalo, J. M. 1998, in ASP Conf. Ser. 142,
        The Stellar Initial Mass Function,
        ed. G. Gilmore \& D. Howell (San Francisco: ASP), 201
\bibitem[Schaller \etal(1992)]{Sch92} Schaller, G., Schaerer, D., Meynet, G.,
        \& Maeder, A. 1992, \aaps, 96, 269
\bibitem[Schmidt-Kaler(1982)]{SK82} Schmidt-Kaler, T. 1982, 
        LB New Series, Vol.\,2b, Astronomy and Astrophysics,
        ed. K. Schaifers \& H. H. Voigt (New York: Springer), 451
\bibitem[Seaquist \etal(1996)]{Sea96} Seaquist, E. R., Carlstrom, J. E., 
        Bryant, P. M., \& Bell, M. B. 1996, \apj, 465, 691
\bibitem[Sellmaier \etal(1996)]{Sel96} Sellmaier, F., Yamamoto, T.,
        Pauldrach, A. W. A., \& Rubin, R. H. 1996, \aap, 305,  L37 
\bibitem[Selman \etal(1999)]{Sel99} Selman, F., Melnick, J., Bosch, G.,
        \& Terlevich, R. 1999, \aap, 347, 532
\bibitem[Shapley \etal(2001)]{Sha01} Shapley, A. E., Steidel, C. C.,
        Adelberger, K. L., Dickinson, M., Giavalisco, M., \& Pettini, M.
        2001, \apj, 562, 95
\bibitem[Shapley \etal(2003)]{Sha03} Shapley, A. E., Steidel, C. C.,
        Pettini, M., \& Adelberger, K. L. 2003, \apj, 588, 65
\bibitem[Shen \& Lo(1995)]{She95} Shen, J. \& Lo, K. Y. 1995, \apj, 445, L99
\bibitem[Shlosman \etal(1989)Shlosman, Frank, \& Begelman]{Shl89} 
        Shlosman, I., Frank, J., \& Begelman, M. C. 1989, \nat, 338, 45
\bibitem[Shopbell \& Bland-Hawthorn(1998)]{Sho98} Shopbell, P. L. 
        \& Bland-Hawthorn, J. 1998, \apj, 493, 129
\bibitem[Sirianni \etal(1998)]{Sir98} Sirianni, M., Nota, A., Leitherer, C.,
        Clampin, M., \& De Marchi, G. 1998, in ASP Conf. Ser. 131,
        Properties of Hot Luminous Stars, ed. I Howarth
        (San Francisco: ASP), 363
\bibitem[Sirianni \etal(2000)]{Sir00} Sirianni, M., Nota, A., Leitherer, C.,
        De Marchi, G., \& Clampin, M. 2000, \apj, 533, 203
\bibitem[Smith \& Gallagher(2001)]{Smi01} Smith, L. J., \& Gallagher, J. S., III
        2001, \mnras, 326, 1027
\bibitem[Sternberg(1998)]{Ste98} Sternberg, A. 1998, \apj, 506, 721
\bibitem[Sundelius \etal(1987)]{Sun87} Sundelius, B., Thomasson, M.,
        Valtonen, M. J., \& Byrd, G. G. 1987, \aap, 174, 67
\bibitem[Telesco \etal(1991)]{Tel91} Telesco, C. M., Campins, H., Joy, M.,
        Dietz, K., \& Decher, R. 1991, \apj, 369, 135
\bibitem[Telesco \& Decher(1988)]{Tel88} Telesco, C. M. \& Decher, R. 
        1988, \apj, 334, 573
\bibitem[Telesco \etal(1993)Telesco, Dressel, \& Wolstencroft]{Tel93}
        Telesco, C. M., Dressel, L. L., \& Wolstencroft, R. D.
        1993, \apj, 414, 120
\bibitem[Telesco \& Gezari(1992)]{Tel92} Telesco, C. M. \& Gezari, D. Y.
	1992, \apj, 395, 461
\bibitem[Thornley \etal(2000)]{Tho00} Thornley, M. D., 
        F\"orster Schreiber, N. M., Lutz, D., Genzel, R., Spoon, H. W. W.,
        Kunze, D., \& Sternberg, A. 2000, \apj, 539, 641
\bibitem[Thronson \& Telesco(1986)]{Thr86} Thronson, H. A., Jr. 
        \& Telesco, C. M. 1986, \apj, 311, 98
\bibitem[Tremonti \etal(2001)]{Tre01} Tremonti, C. A., Calzetti, D.,
        Leitherer, C., \& Heckman, T. M. 2001, \apj, 555, 322
\bibitem[Vanzi \etal(1998)Vanzi, Alonso-Herrero, \& Rieke]{Van98}
        Vanzi, L., Alonso-Herrero, A., \& Rieke, G. H. 1998, \apj, 504, 93
\bibitem[Wamsteker(1981)]{Wam81} Wamsteker, W. 1981, \aap, 97, 329
\bibitem[Wei{\ss} \etal(1999)]{Wei99} Wei{\ss}, A., Walter, F., Neininger, N.,
        \& Klein, U. 1999, \aap, 345, L23
\bibitem[Weitzel \etal(1996)]{Wei96} Weitzel, L., Krabbe, A., Kroker, H.,
        Thatte, N., Tacconi-Garman, L. E., Cameron, M., \& Genzel, R. 
	1996, \aaps, 119, 531
\bibitem[Wild \etal(1992)]{Wil92} Wild, W., Harris, A. I., Eckart, A.,
        Genzel, R., Graf, U. U., Jackson, J. M., Russell, A. P. J.,
	\& Stutzki, J. 1992, \aap, 265, 447 
\bibitem[Wright \etal(1988)]{Wri88} Wright, G. S., Joseph, R. D.,
        Robertson, N. A., James, P. A., \& Meikle, W. P. S.
        1988, \mnras, 233, 1
\bibitem[Yun \etal(1993)Yun, Ho, \& Lo]{Yun93} 
        Yun, M. S., Ho, P. T. P., \& Lo, K. Y. 1993, \apj, 411, L17
\bibitem[Yun \etal(1994)Yun, Ho, \& Lo]{Yun94} 
        Yun, M. S., Ho, P. T. P., \& Lo, K. Y. 1994, \nat, 372, 530
\bibitem[Zhou \etal(1984)]{Zho84} Zhou, K., Hao, Y., Chen, P., Zhang, Y.,
        \& Gao, M. 1984, Ap{\&}SS, 107, 373

\end{thebibliography}
\end{document}